\documentclass{rAMF2e}
\usepackage{epstopdf}
\usepackage{subfigure}

\usepackage{verbatim}
\usepackage{url}
\usepackage[utf8]{inputenc}

\begin{document}
\articletype{ARTICLE}

\title{Sharper asset ranking from total drawdown durations}

\author{
\name{Damien Challet$^{\ast,\dag}$\thanks{{\em{Correspondence Address}}: D. Challet, Laboratoire MICS, CentraleSupélec, Université Paris-Saclay, 92290 Châtenay-Malabry, France. Email: damien.challet@centralesupelec.fr}}
\affil{$^\ast$Laboratoire MICS, CentraleSupélec, Université Paris-Saclay, France, $^\dag$Encelade Capital SA, Lausanne, Switzerland}
\received{9 November 2015}
}
\maketitle

\begin{abstract}
The total duration of drawdowns is shown to provide a moment-free, unbiased, efficient and robust estimator of Sharpe ratios both for Gaussian and heavy-tailed price returns.  We then use this quantity to infer an analytic expression of the bias of moment-based Sharpe ratio estimators as a function of the return distribution tail exponent. The heterogeneity of tail exponents at any given time among assets implies that our new method yields significantly different asset rankings than those of moment-based methods, especially in periods large volatility. This is fully confirmed by using 20 years of historical data on 3449 liquid US equities.
\end{abstract}

\begin{keywords}
Asset ranking, Sharpe ratio, drawdowns, unbiased estimator, heavy tails
\end{keywords}

\section{Introduction}

Sharpe ratios \citep{sharpe1964capital} appear naturally in financial
analysis for a good reason: they are nothing else than signal-to-noise
ratios, a fundamental quantity in signal analysis. In a Gaussian world,
they are also equivalent to the t-statistics. Finance is not
an ideal world, however, and many problems arise in practice. Sharpe
ratio's distribution \citep{lo2002sharpe}, bias \citep{miller1978sample,jobson1981performance}
and corrections due to serial correlations  \citep{lo2002sharpe,mertens2002lo,christie2005sharpe,opdyke2007comparing} have
been characterized. Better estimating methods use the Generalized
Moments Method \citep{lo2002sharpe,christie2005sharpe} and block
bootstraps \citep{ledoit2008robust}. Although Sharpe ratios only
depend on the first and second moments of price returns, their
variance depends on the third and fourth moments \citep{lo2002sharpe,mertens2002lo,christie2005sharpe,opdyke2007comparing}.
Given the definition of the Sharpe ratios, it is not surprising that
all these methods rely on the computation of moments of price returns.
But as noted e.g. in \citet{opdyke2007comparing}, this may be problematic
as the fourth moment may not be defined \citep{daco,jondeau2003conditional}. Finally, the standard estimator of the Sharpe ratio is known be biased for heavy-tailed price returns. Once again, the corrections proposed e.g. for the Deflated Sharpe Ratio, depend on the third and fourth moment \citep{bailey2014deflated}.

Here, I propose a new way to estimate Sharpe ratios that does not require the computation of any moment and that may be extended to estimate the drift of time series with infinite variance. It is based on the fact that the total duration of all drawdowns in a price time series of a given length is a monotonic function of the Sharpe ratio; by symmetry, the same holds for the total duration of all drawups. As a consequence, one may estimate Sharpe ratios by computing the difference between the total durations of drawups and drawdowns. This quantity is bounded by definition and leads to an estimator that is both robust to outliers and more efficient than direct estimates of Sharpe ratios for heavy-tailed data.

Above all, the new estimator is unbiased for heavy-tailed data, in contrast to the standard estimation method. Even more, we propose that the Sharpe ratio depends in a simple way of total drawdown durations and return distribution tail exponent, which allows a direct estimation of the bias of the moment-based  estimator at fixed total drawdown duration length. This gives a new take on asset ranking: because all the asset price return distributions have different tail exponents at any point in time, the new method yields considerably different asset rankings, especially at the top and bottom quantiles, and during the most volatile periods. Thus this paper contributes both theoretically and empirically to the on-going debate about the relevance of the ranking method  \citep{eling2007does,zakamulin2010choice,schuhmacher2011sufficient,ornelas2012yes,auer2013performance,auer2013robust}

\begin{figure}
\centerline{\includegraphics[width=0.6\textwidth]{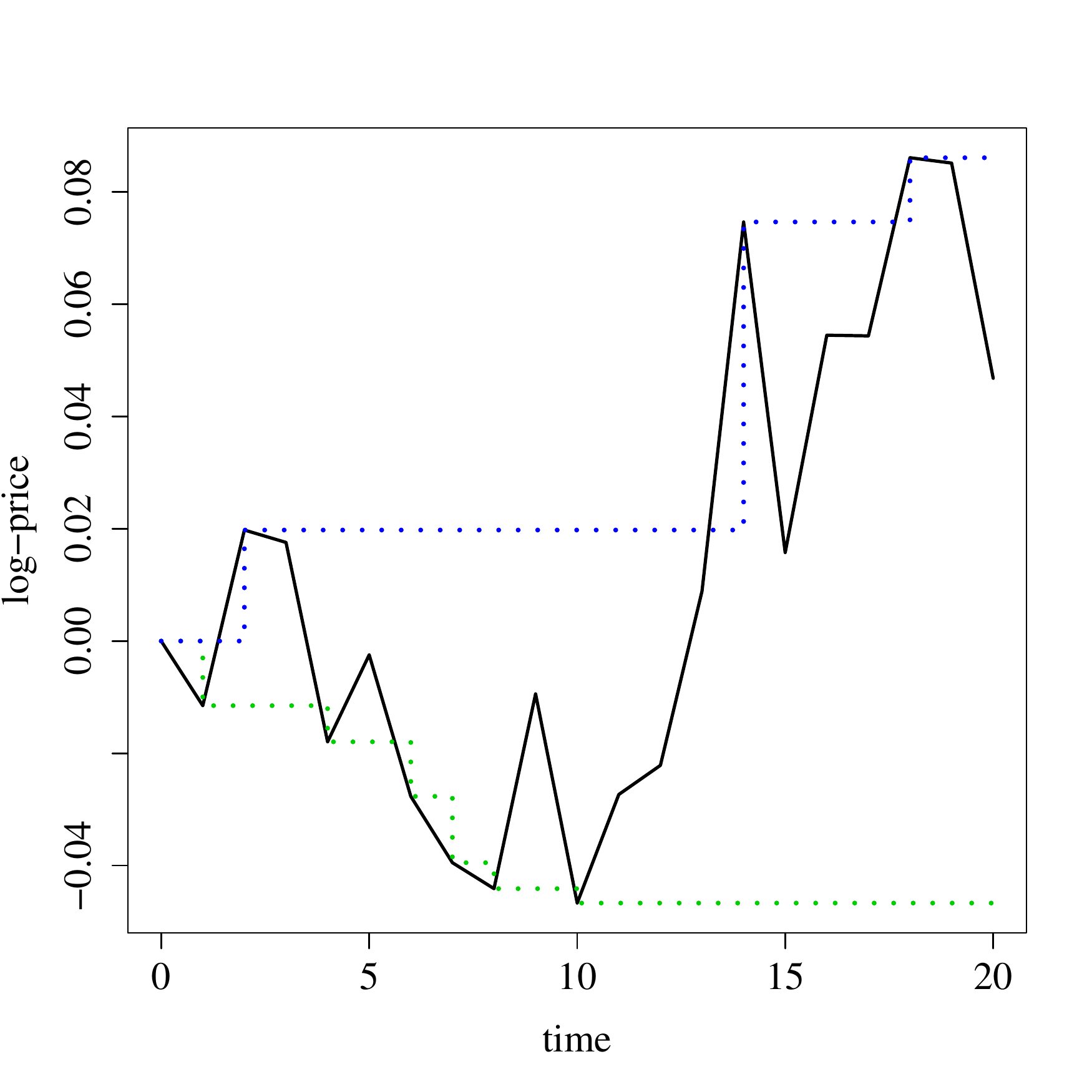}}
\protect\caption{Example log price time series (black lines), its running maximum (blue dashed
lines), and running minimum (green dashed lines). The number of upper (lower)
records $R_+$  ($R_-$) is equal to the number of jumps of the running maximum (minimum)
plus one since the first point counts as a record by convention: here
$R_{+}=4$ and $R_{-}=7$. 
 The total drawdown duration is $T_{-}=16$
and the total drawup duration is $T_{+}=13$. Clearly, $R_{+}+T_{-}=R_{-}+T_{+}=20+1$,
the number of returns plus one.\label{fig:Example-price-time}}
\end{figure}

Intuitively, the sum of all drawdown durations, i.e., the total
drawdown duration of a time series of fixed length, is linked to the
number of upper price records since a new price return pushes the
price either to an all time high (a new upper record) or to a drawdown
(see Fig.~\ref{fig:Example-price-time}). This implies that if $n$ is
the length of a price time series and $R_+$ is the number of its upper 
records ($R_+\ge1$ because the first point is a record by convention), the total drawdown duration, denoted by $T_-$, is
$T_-=n-(R_+-1)$. Because of this equivalence, total drawdown/up duration
and the numbers of price records lead to two equivalent estimators;
accordingly, we will use either wordings.  Assuming that log prices
are simple random walks, drawdown/up durations are determined by
first-passage times, themselves derived from persistence (or survival)
properties \citep{redner2001guide}. The connection between persistence
and price dynamics, especially in the context of market
microstructure, is well known
\citep{lo2002econometric,eisler2009diffusive}.

Persistence is at the core of a noteworthy recent result about
discrete-time unbiased random walks. In a financial context, it may be
stated as follows: the distribution of the number of upper (or lower)
records of a price time series with independent and identically
distributed return (i.i.d.), of a fixed length, does not depend on the
increment distribution provided that the latter is symmetric and
continuous \citep{majumdar2008universal}. This universality is behind
the robustness and power of the r-statistics, a family of statistics
based on the number of records of a time series, which not only
provides a powerful non-parametric location test
\citep{challet2015rstatistics} but also, as shown here, an efficient
estimator of Sharpe ratios.  Their robustness come from the fact that
the influence of outliers is much dampened because sample values are
transformed into an integer number with bounded admissible values.

\citet{majumdar2012record} show that the distribution of the number of
records converges to a Gaussian distribution in the limit of
infinitely long time series provided that the price return
distribution has a finite variance. Even better, the support of the
finite-size sample distribution of the new estimator is bounded,
contrarily to that of Sharpe ratios (and t-statistics), and is
accordingly more peaked than a normal distribution
\citep{challet2015rstatistics}. When the true Sharpe ratio is
different from zero, the expected number of records and its variance
are distribution-dependent; exact expressions are only known for
exponentially distributed increments, hence one has to resort to
approximations and numerical simulations for other types of
distribution in the limits of large and small Sharpe ratios.

Drawdown durations are by definition integer numbers, which is
not optimal to estimate a real number. The solution comes from random
permutations. Assuming that the price returns are i.i.d., one can
shuffle their order at will and compute the resulting price time series,
which is an equally valid representation of a given set of price returns and most likely have a different number of upper and lower records. Thus, to obtain a more precise estimate of
the Sharpe ratio, one takes the average of the difference between the total drawdown and drawup durations over many such permutations (see Fig.\ \ref{fig:R0bar} for a graphical explanation).

The structure of this paper is as follows: Section 2 introduces the
necessary notations to define price record statistics and shows that
when prices have a positive trend, heavy-tailed increments lead to a
larger number of upper price records than Gaussian increments; a
mathematical derivation of the expected number of price records for
Student's t-distributed increments is reported in Appendix A, which
focuses on the case of tail exponent equal to 4 (3 degrees of freedom)
for the sake of analytical tractability. Section 3 investigates the
efficiency of the number of price records as Sharpe ratio estimators
relative to the vanilla estimator and shows that the new estimator is
several times more efficient than moment-based methods for
heavy-tailed variables and almost as efficient as the vanilla
estimator in the case of Gaussian variables; it then derives a simple
equation that simplifies the calibration of the relationship between
true Sharpe ratio and estimated tail exponents and number of records.
Section 4 uses an unbiased historical data set of 3449 liquid US
equities to estimate 100-day rolling Sharpe ratios with both
methods. It turns out that in leptokurtic times, the estimates from
both methods may differ very significantly because the vanilla Sharpe
ratio estimator is not only more volatile, but also systematically
overestimates the information content of price time series that have
heavy-tailed returns.

\section{Record statistics of random prices}

Financial data exist in discrete time, which will be the point of view
adopted in this paper. Let us assume that the initial log price is
$S_{0}=0$ and that its value at time $k>0$ follows
\begin{equation}
S_{k}=S_{k-1}+r_{k}+c
\label{eq:price}
\end{equation}

 where $r_{k}$ is the increment at time $k$, assumed to be identically
 and independently drawn from a continuous distribution $P(r)$, and
 $c$ is a constant trend.  Let the running maximum $M_{k}=\max_{1\le
   t\le k}S_{t}$ (see Fig.\ \ref{fig:Example-price-time}).  The number
 of upper records of a time series of length $n$ is the number of
 jumps of $M_{n}$, which by convention always includes $M_{1}$; it
 will be denoted by $R_{+}$ and its distribution by $P(R_{+},n)$. In
 the same spirit, one defines $R_{-}$, the number of lower records, as
 the number of jumps of the running minimum.

\citet{majumdar2008universal,ledoussal2009driven,majumdar2012record}
demonstrate that many quantities of interest are fully characterized
by the persistence function $q_{-}(n)$ of the process, i.e., the
probability that the price has never exceeded its starting value after
$n$ steps. It is advantageous to work with its characteristic function $\tilde{q}_-(z)=\sum_{n>0} z^nq_-(n)$. 

For example, the characteristic function of $P(R_{+},n)$
is \citep{majumdar2008universal}
$$\tilde{P}(R_{+},z)=\tilde{q}_{-}(z)[1-(1-z)\tilde{q}_{-}(z)]^{R_+-1},$$
 while the characteristic function
of the expected number of upper records $m_{+}(n)=E(R_{+})(n)$ can
be written as $\tilde{m}_{+}(z)=[(1-z)^{2}\tilde{q}_{-}(z)]^{-1}$
\citep{ledoussal2009driven}. 

Generalized Sparre Andersen theorem
\citep{andersen1953fluctuations,feller2008introduction} provides a
constructive way to compute $\tilde{q}_{-}$: for any continuous and
symmetric $P(r)$,
\begin{equation}
\log\left(\tilde{q}_{-}(z)\right)=\sum\limits _{n=1}^{\infty}\frac{z^{n}}{n}P(S_{n}<0).\label{eq:S-A}
\end{equation}

\subsection{Driftless prices}

A direct consequence of this theorem is the universality of the
unbiased case $c=0$ since $P(S_{k}<0)=\frac{1}{2}$ for all symmetric
and continuous distributions, as indeed
$\tilde{q}_{\pm}(z)=\tilde{q}(z)$ is the same for all such
distributions and

$$P(R,n)={2n-R+1 \choose R}2^{-2n+R-1},$$ where $R$ may either be
$R_{+}$ or $R_{-}$, by symmetry \citep{majumdar2008universal}.  This
implies that the first two moments of this distribution are
$$
E(R_\pm)(n)\simeq 2\sqrt{\frac{n}{\pi}},~~~\textrm{and}~~~~E[((R_\pm-E(R_\pm))^2](n)\simeq (2-4/\pi)n.
$$

\subsection{Limit of small relative drift}

Analytical results are harder to obtain in the case of non-zero drift
($c\ne 0$) since Sparre Andersen theorem requires the full knowledge
of all convolutions of the elementary increments. Denoting the
standard deviation of the increments $r_k$ by $\sigma$, good
approximations of the expected number of upper records are known for
Gaussian increments in the limit of small relative drift, i.e., when
$c/\sigma\ll1$ and $n\gg1$ while $cn/\sigma\ll 1$
\citep{wergen2011record,majumdar2012record}:
\begin{equation}\label{eq:mn_Gauss}
E(R_+)(c/\sigma,n)\simeq 2\sqrt{\frac{n}{\pi}}+\frac{c\sqrt{2}}{\sigma\pi}\left[n\arctan(\sqrt{n})-\sqrt{n}\right].
\end{equation}

The case of heavy-tailed increments with finite variance has not been
thoroughly investigated. We will focus on Student's t-distributions
because of their abilities to reproduce both fat-tailed and Gaussian
returns. They are known to describe the unconditional price return
distribution (i.e., forgetting about volatility heteroskedasticity)
\citep{bouchaudpotters,longin2005choice,opdyke2007comparing} and
innovations (see e.g. \cite{bollerslev1987conditionally}). Let us
therefore assume from now on that the price returns $r_{k}$ are
distributed according to a Student's t-distribution of variance
$\sigma^2$ with $\nu$ degrees of freedom (we use this wording only to
parametrize the return distribution), denoted by $P(r)$. Sparre
Andersen theorem requires the knowledge of the $n$-time convoluted
return distribution, denoted by $P^{(n)}(r)$, of which no explicit
expression exists for generic values of $n$ and $\nu$. In passing,
$P^{(n)}(r)$ can be explicitly computed for any value of $n$ provided
that $\nu$ is odd but the expressions quickly become cumbersome as $n$ grows \citep{nadarajah2005convolutions}. This is why we shall resort to approximations.

\begin{figure}
\centerline{\includegraphics[width=0.6\textwidth]{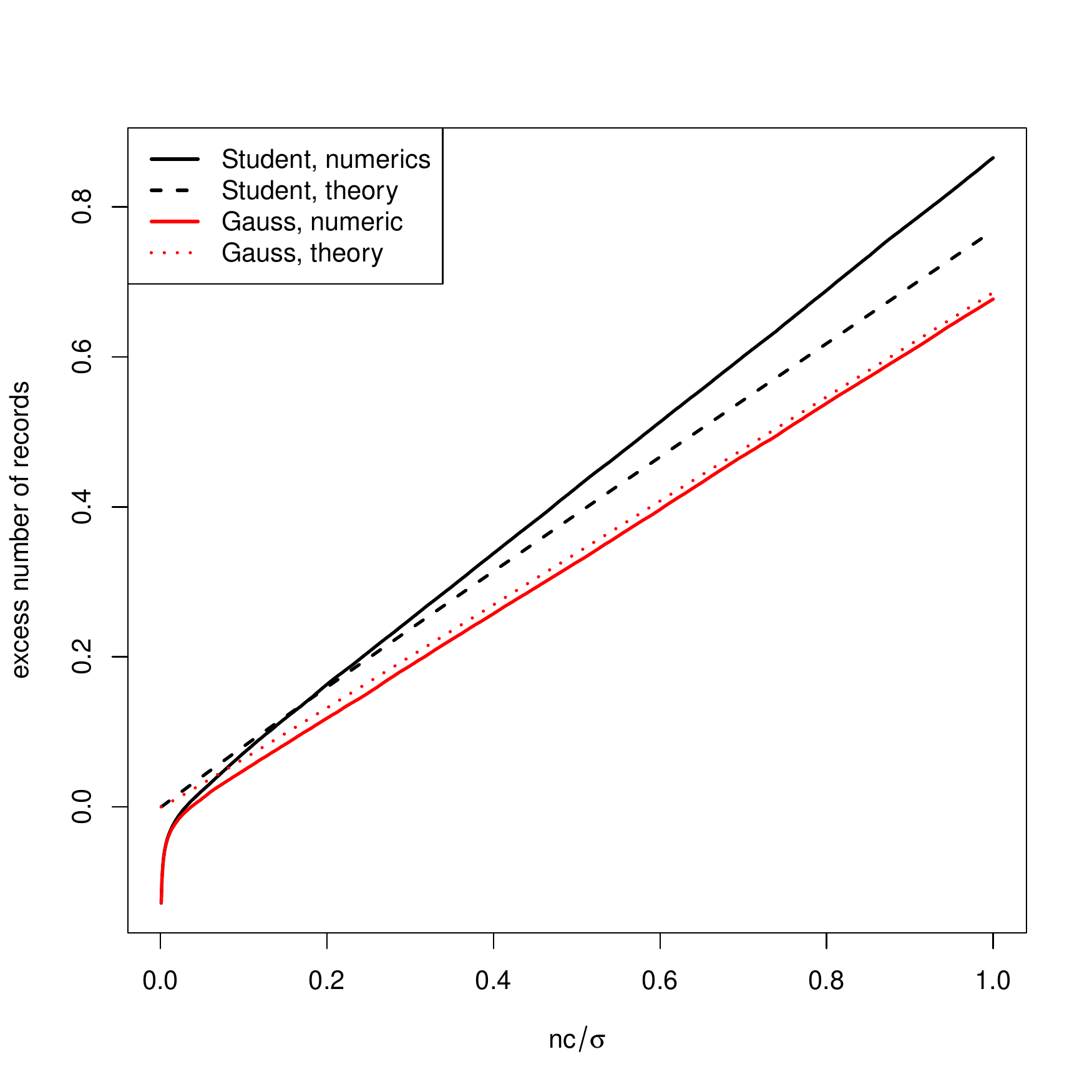}}\protect\caption{Excess number of records $E(R_+|c/\sigma,n)-E(R_+|0,n)$ for biased random walks with Student-t increments
($\nu=3$). Interrupted lines are theoretical predictions and continuous
lines are from numerical simulations. $c=0.001$, $\sigma=1$, averages
over $10^{7}$ samples.\label{fig:excess}}
\end{figure}

Appendix A reports approximate analytical results for the case $\nu=3$, i.e., for a tail exponent of 4.\footnote{This precise value is the only one for which analytical  computations seem workable. It also happens to be in
line with the average tail exponent of US equities daily and intraday
price returns  \citep{jansen1991frequency,plerou1999scaling,bouchaudpotters,longin2005choice}.} The resulting expected number of upper records becomes, in the same limit   $c/\sigma\ll1$ and $n\gg1$ while $cn/\sigma\ll 1$,
\begin{eqnarray}
E(R_+)(n,c/\sigma)&\simeq&\frac{2\sqrt{n}}{\sqrt{\pi}}+\frac{c\sqrt{2}}{\sigma\pi}\left[n\arctan(\sqrt{n})-\sqrt{n}\right]\nonumber\\&&+\frac{c}{\sigma}\frac{8}{\sqrt{3}\pi^{3/2}}\sqrt{n}\left(\mbox{atanh}\sqrt{1-\frac{1}{n}}-\sqrt{1-\frac{1}{n}}\right).\label{eq:m(n)_student_main}
\end{eqnarray}
Although a first order expansion, Eq.\ (\ref{eq:m(n)_student_main}) is not very accurate even in
the limit of small $n(c/\sigma)$, because the approximations needed to obtain explicit equations are
quite rough (see Fig.\ \ref{fig:excess}).
 However, it was worth computing it for several reasons. First, it contains the correct dependence of $E(R_+)(n,c/\sigma)$ on $n$ for small Sharpe ratios, which means that one may use this functional form to fit numerical simulations. 
Second, the presence of the third term, due to the difference between Gaussian and t-distributions at the origin, correctly implies that the prices with positive trends and heavy tails (and small Sharpe ratios) have a larger expected number of price records, which emphasises the importance of accounting for the tails of price return distributions when using price records to estimate Sharpe ratios (cf. section \ref{sub:estimate}). 
 
Appendix B contains the derivation of   $E(R_+)(n,c/\sigma)$ in the large relative drift limit, i.e., $c/\sigma\gg1$ and $n\gg1$. In this case, the expected number of records grows linearly.

It is noteworthy that these limits do not unequivocally correspond to small and large Sharpe ratios, since the limits do not involve $n^{1/2}$ but $n$. The small effective drift limit can be rewritten as $c/\sigma\sqrt{n}\ll1/\sqrt{n}$, which correspond to vanishingly small Sharpe ratios, of little relevance to Finance; there is no guarantee that the very large $c/\sigma$ limit corresponds to realistic situations. Thus, depending on both $c/\sigma$ and $n$, one may be close to either limit, or in  a no limit's land. As a consequence, the next section resorts to extensive numerical calibration.

\section{Moment-free Sharpe ratio estimator}
\label{sub:estimate}
\begin{figure}
\centerline{\includegraphics[width=0.85\textwidth]{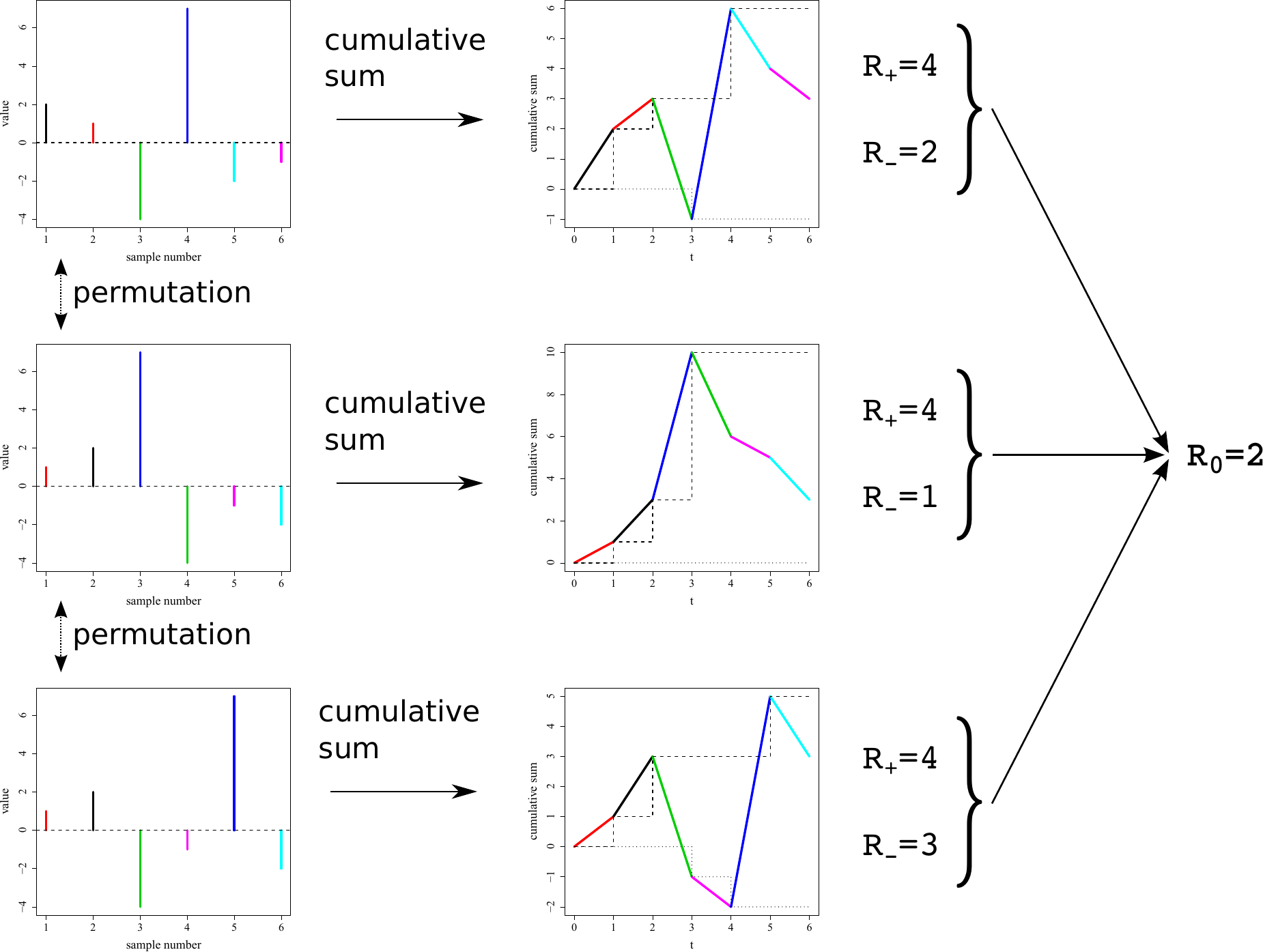}}
\protect\caption{Schematic explanation of the idea behind the permutation estimator of Sharpe ratios: one computes the difference between total drawup and drawdown durations, or equivalently, the number of jumps of the running maximum (dashed lines) and the number of jumps of the running minimum (dotted lines) of the cumulated sums of the sample values, averaged over many random permutations of the price returns. By convention, the first point counts as a first record for both the running maximum and minimum.\label{fig:R0bar}}
\end{figure}

\label{sub:Single-asset-Sharpe-ratio}

As shown by \cite{majumdar2012record}, the expected number of records is a monotonous function of the ratio $c/\sigma$, hence, of the Sharpe ratio. In other words, there is a one-to-one correspondence between the two quantities. This implies that it is possible to estimate Sharpe ratios from the number of upper or lower records.  More precisely, we have
\begin{equation}\label{eq:E Rplus SR}
E(R_+)=F_+(c/\sigma),
\end{equation}
 which implies that 
\begin{equation}
\hat{\left(\frac{c}{\sigma} \right)}=F_+^{-1}(\hat {R}_+).
\end{equation}
Because of the lack of exact results, we shall use numerical simulations to calibrate $F$.

The main problem of a number of records is that it is an integer number
by definition, which yields an estimator with unacceptable
precision for short time series. The fundamental idea of the r-statistics \citep{challet2015rstatistics}, in this context,
consists in assuming that its
log returns are i.i.d.. In that case, one may build many other
log price paths based on random permutations of the original returns and
thus measure the average number of records of the cumulated sums
over many permutations (see Fig.\ \ref{fig:R0bar}) (this scheme may be extended to correlated time series). Mathematically, denoting the random permutation
of index $i\in\{1,\cdots,n\}$ by $\pi(i)$ and the ensemble of all
permutations by $\Pi$, the average number of records is $\bar{R}_{+}=\frac{1}{|\Pi|}\sum_{\pi\in\Pi}R_{+,\pi}$
where $R_{+,\pi}$ is the number of upper records of $S_{n,\pi}=\sum_{m=1}^{n}r_{\pi(m)}$.
In practice, one restricts computations to a subset of $\Pi$ for the sake of computational tractability, which has little influence on the end result; in this study, we have used 1000 random permutations.  The new Sharpe ratio estimator is then based on $R_{0}=\bar{R}_{+}-\bar{R}_{-}$. More precisely, the idea is to first calibrate the relationship $E(R_0)=F_0(c/\sigma,n)$ at fixed $c/\sigma$ for a given distribution of synthetic price returns. Denoting $\theta=c/\sigma$, one then inverts this relationship to obtain 

\begin{equation}\label{eq:theta_from_R0}
\hat \theta= F^{-1}(\hat R_0,n).
\end{equation}

Estimation in the rest of this paper is based on extensive numerical simulations to establish the relationships $E(R_{0})$ as a function of parameters $n$, $\theta$, and the Student parameter $\nu$. We chose $\nu=\{2.5,\cdots,10\}$ with increments of 0.1, $10<n<375$ by steps of 5 and $n=504$; we take 31 values of $\theta\in[0.001,1]$ growing according to a geometric series. For each triple $(n,\theta,\nu)$, we generate $N_{avg}$  synthetic time series, estimate $R_0$ over 1000 random permutations for each time series and then average $R_0$ over the $N_{avg}$ time series. Splines are then used to fit and invert the relationships of Eq.\ (\ref{eq:theta_from_R0}).

\begin{figure}
\centering{\includegraphics[width=0.5\textwidth]{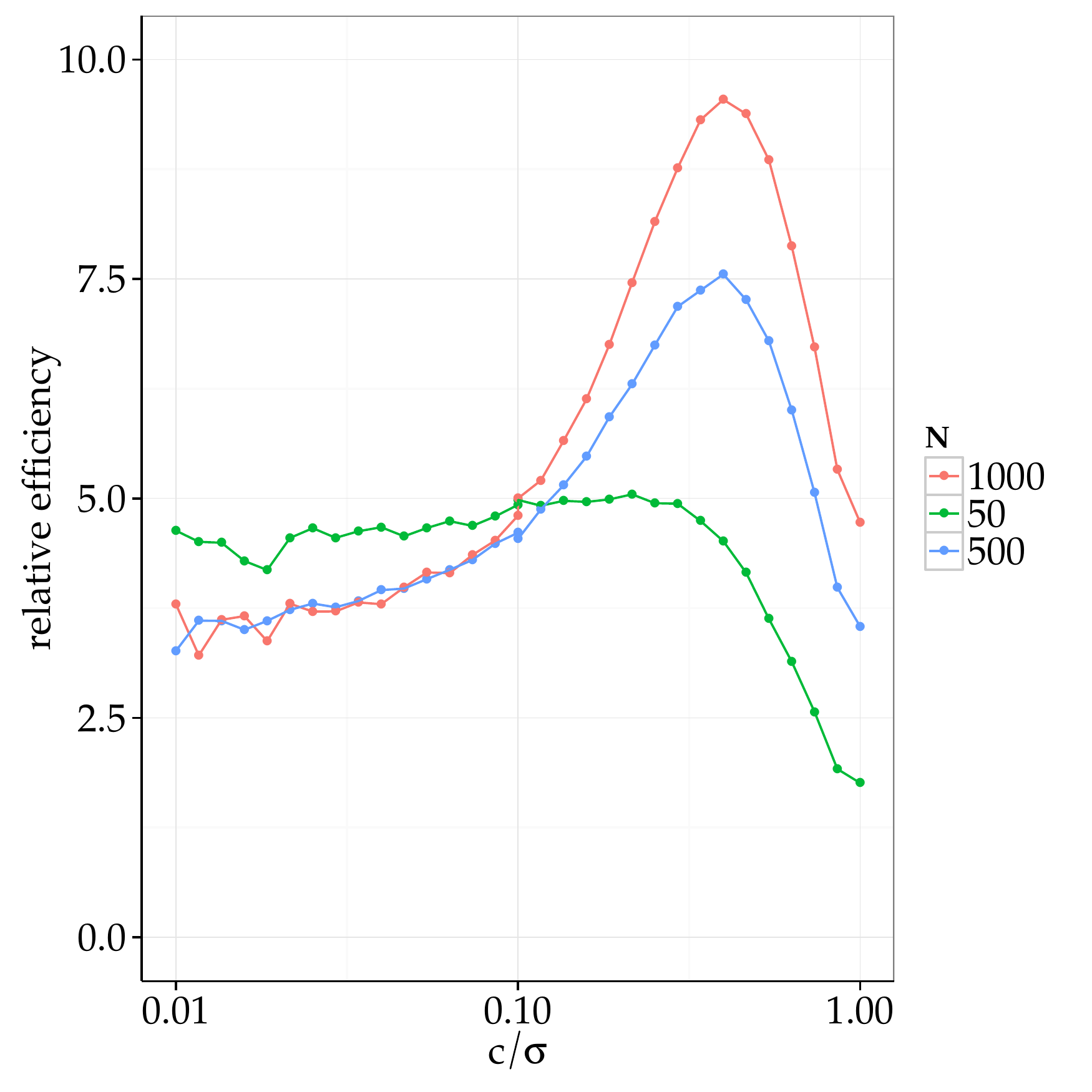}\includegraphics[width=0.5\textwidth]{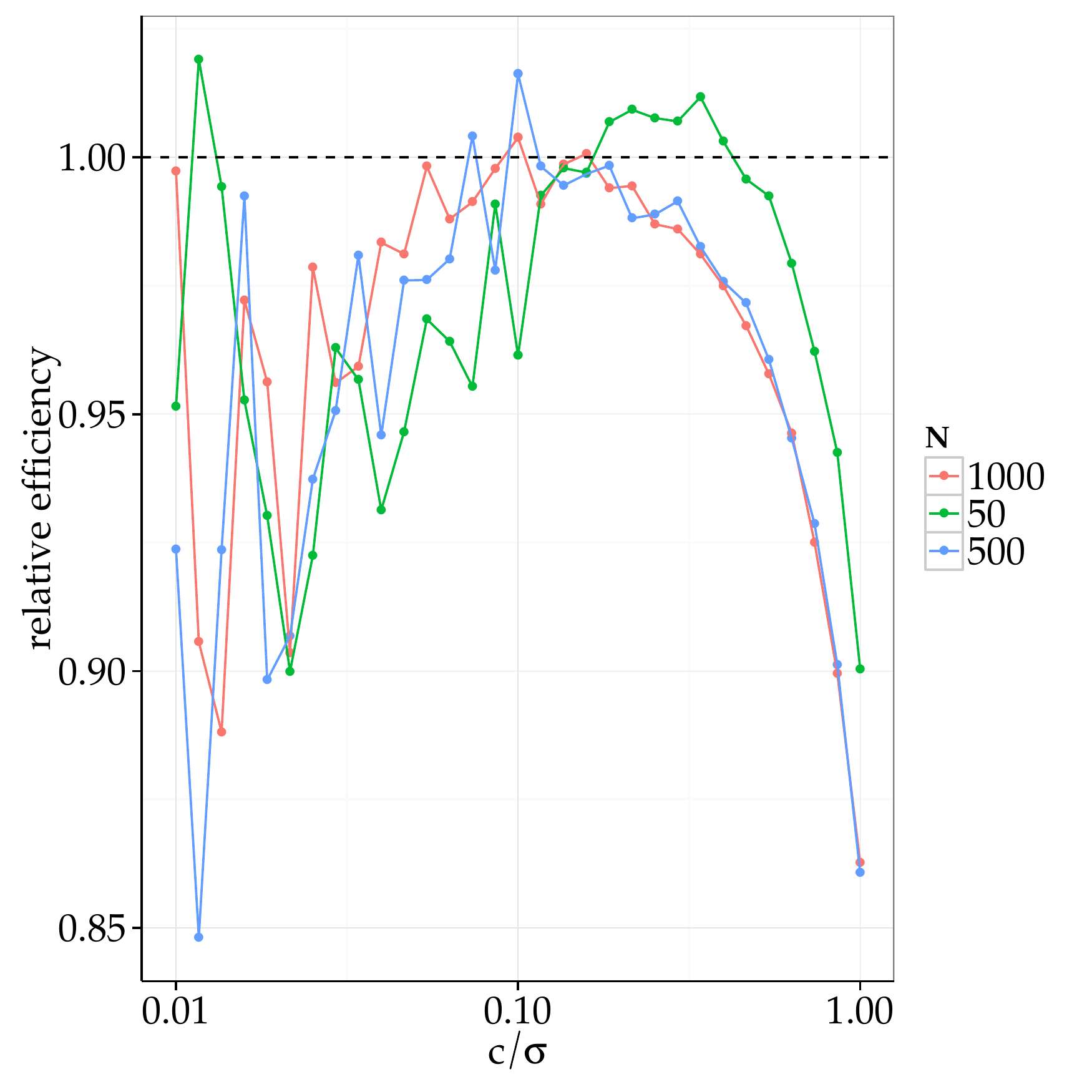}}
\protect\caption{Efficiency of the record-based estimator $\theta_{0}$ relative to
that of the vanilla estimator, defined by the ratio of
the variance of the new estimator $\theta_{0}$ and the usual one
$\theta_{S}$ as a function of the true Sharpe ratio $c/\sigma$ of
the synthetic data. Averages over $N_{avg}=10^{6}$ samples per point; record numbers have been 
averaged over $1000$ permutations; left plot: Student-distributed increments
with tail exponent set to 4; right plot: Gaussian increments.\label{fig:efficiency_theta_StGa}
}
\end{figure}

\subsection{Efficiency}

Moment-based estimators have a hard time with heavy-tailed data. It is thus clear that their precision, i.e., efficiency, suffer from heavy tails. The new estimator, on the other hand, is likely to be less affected by the latter.

In order to compare their respective efficiency, let us denote by $\theta_{0}$ the Sharpe ratio inferred from $R_{0}.$ The standard deviation of $\theta_{0}$, denoted by $\sigma_{\theta}$, is obtained by the method of Deltas, i.e.,
from the relationship $\sigma_{\theta}=\sigma_{R}\frac{1}{\frac{dE(R_{0}|n,\theta)}{d\theta}}$ where $\sigma_{R}$ is the standard deviation of $R_{0}$; the numeric
derivative of $E(R_{0}|n,\theta)$ was computed numerically with splines.
The relative efficiency of $\theta_{0}$ with respect to the straightforward
estimator $\theta_{S}=\hat{\mu}/\hat{\sigma}$ is then defined as
$\rho=\sigma_{S}^{2}/\sigma_{R}^{2}$ where $\sigma_{S}$ is the standard
deviation of $\theta_{S}$. Left plot of Fig.\ \ref{fig:efficiency_theta_StGa} reports the relative efficiency of $\theta_{0}$ for various $n$ for Student's t-distributed returns and  $\nu=4$. The new estimator is unambiguously more powerful than the vanilla estimator. This result holds as long as the returns are heavy tailed.

Financial price returns are not heavy tailed all the time. Thus it is important to check the efficiency of record
statistics for log prices with Gaussian increments. Since the vanilla estimator is asymptotically optimal in this case \citep{neyman1933problems}, any other estimator is bound
to be less efficient for large $n$. The right hand side plot of Fig.\ \ref{fig:efficiency_theta_StGa} plots the relative efficiency of $\theta_{0}$ for Gaussian increments,
which depends on $c/\sigma$. Remarkably, $\theta_{0}$ may be  slightly  more
efficient than the t-statistics itself  for small $n$.

\subsection{Dependence on Student tail exponent}

\begin{figure}
\centering{}\includegraphics[width=0.6\textwidth]{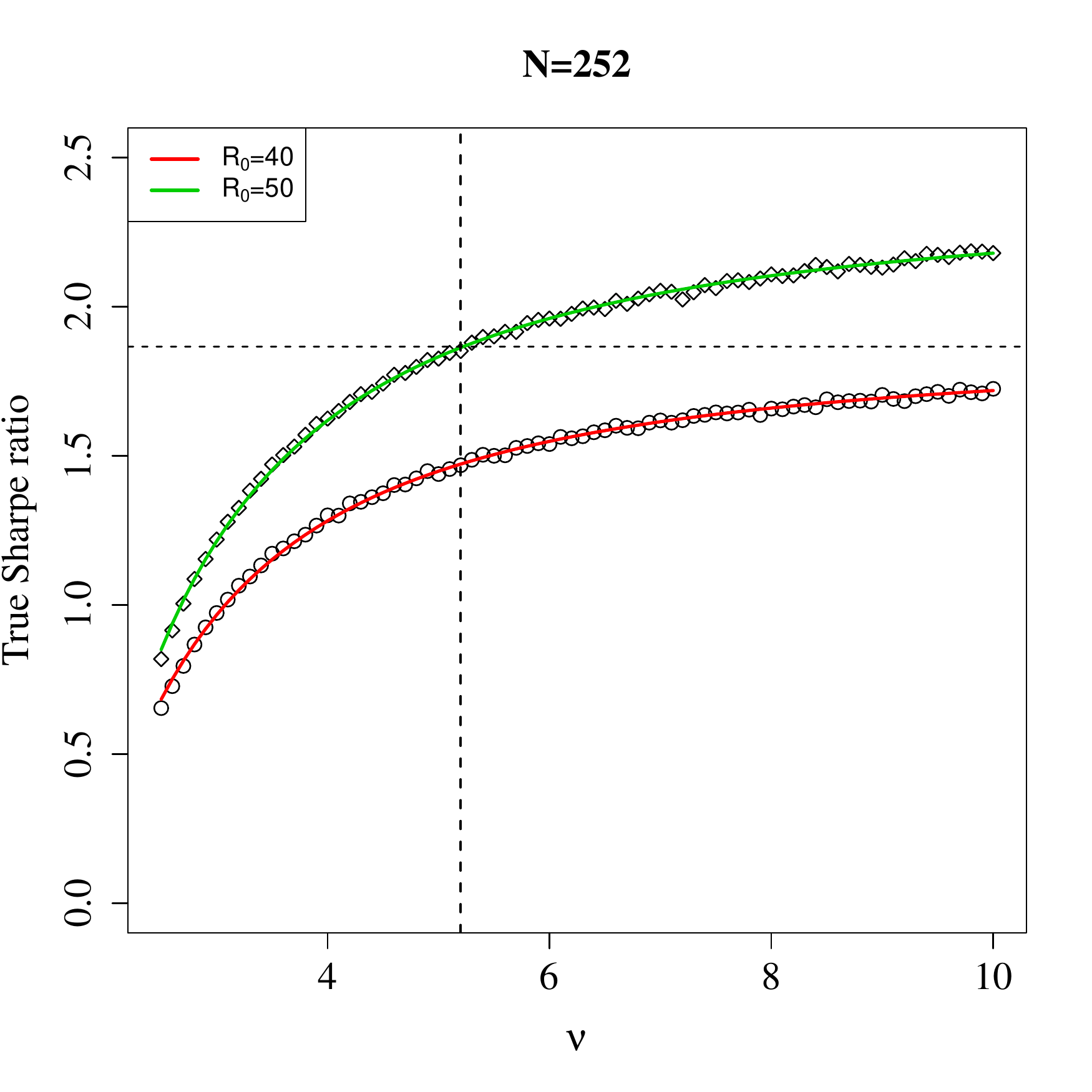}\protect\caption{True Sharpe ratio versus Student tail index $\nu$ for $R_0=40$ and $50$ (circles and lozanges, respectively) for time series of length 252. The continuous lines are least-squares regressions to Eq. (\ref{eq:fitSR_vs_nu}). The horizontal dashed line reports the value of the true Sharpe ratio for Gaussian increments ($\nu\to\infty$) and $R_0=40$, while the vertical one  stands at $\nu=5.2$, which is the threshold at which the respective rank switches if a time series with $R_0=40$ and $\nu=\infty$ is compared with one with $R_0=50$ and with exponent $\nu$.\label{fig:SR_vs_nu}}
\end{figure}

Although only the $\nu=3$ was studied analytically above, as it leads to workable expressions, the relationship between $E(R_0)$ and the Sharpe ratio of increments with a Student's t-distribution depends on $\nu$. As a consequence, at fixed $n$ and $R_0$, the estimated Sharpe ratio also depends on $\nu$. Extensive numerical simulations  (see Fig.\ \ref{fig:SR_vs_nu}) with $N_{avg}=10^5$ show that, at fixed $R_0$ and $n$,
\begin{equation}
E_\nu(\hat{\theta})=a(R_0,n)-b(R_0,n)\nu^{-3/2},\label{eq:fitSR_vs_nu}
\end{equation}
where $a(R_0,n)=E_{\infty}(\hat{\theta})$ corresponds by definition to the average (and unbiased) Sharpe ratio of a process with Gaussian increments  ($\nu\to\infty$). 

In addition to providing a simple way to extend the inference of the Sharpe ratio for arbitrary large values of $\nu$ from a finite interval of $\nu$, this equation quantifies the bias of an estimation of Sharpe ratios if one neglects the effect of non-Gaussian returns. Indeed, the estimation of $R_0$ does not require any assumption about the underlying increment distribution, only the connection to the Sharpe ratio does. This means that estimating Sharpe ratios requires estimating $\nu$, and that assuming $\nu=\infty$, as the vanilla method does,  overestimates the true value of the Sharpe ratio as soon as the price return  distribution has fatter tails than a Gaussian one. As a consequence, results about the equivalence of ranking from various methods only hold if all the assets have the same tail exponent. This point is further discussed in section \ref{ranking}.

\subsection{Simplified estimation}

Equation (\ref{eq:fitSR_vs_nu}) provides a great complexity reduction in the estimation of the Sharpe ratio, but estimation may be further simplified by studying the dependence of both $a$ and $b$ on $R_0$ and $n$. We choose values of $R_0$ from 1 to $n$, and inferred the corresponding values of $\theta$ thanks to the calibrated relationships of Eq.\  (\ref{eq:theta_from_R0}). Then, at fixed $n$, we choose a value $R_0$ and perform the non-linear fit of Eq.\ (\ref{eq:fitSR_vs_nu}), which yields $a(R_0,n)$ and $b(R_0,n)$. We then filter out fits whose  p-value associated with $b$ is larger than 0.01 and whose average square residual is larger than 0.1, which only happens in regions with large $n$ and small $R_0$ (in other words, where $\hat\theta=0$ is a fairly good approximation of the true Sharpe ratio). This leaves 13282  values of $a$ and $b$, one for each remaining couple $(R_0,n$).

\begin{figure}
\centerline{\includegraphics[width=0.48\textwidth]{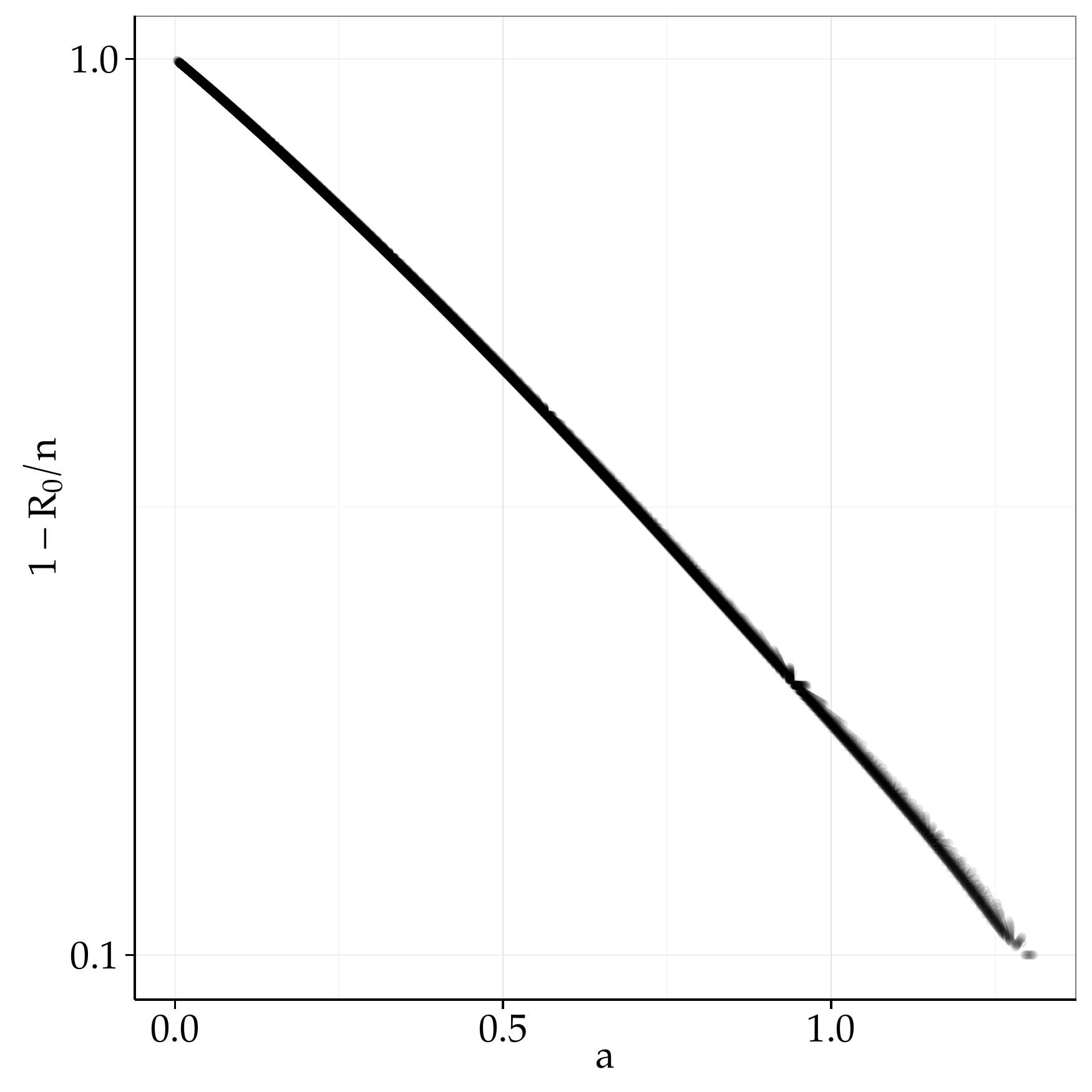}~~\includegraphics[width=0.48\textwidth]{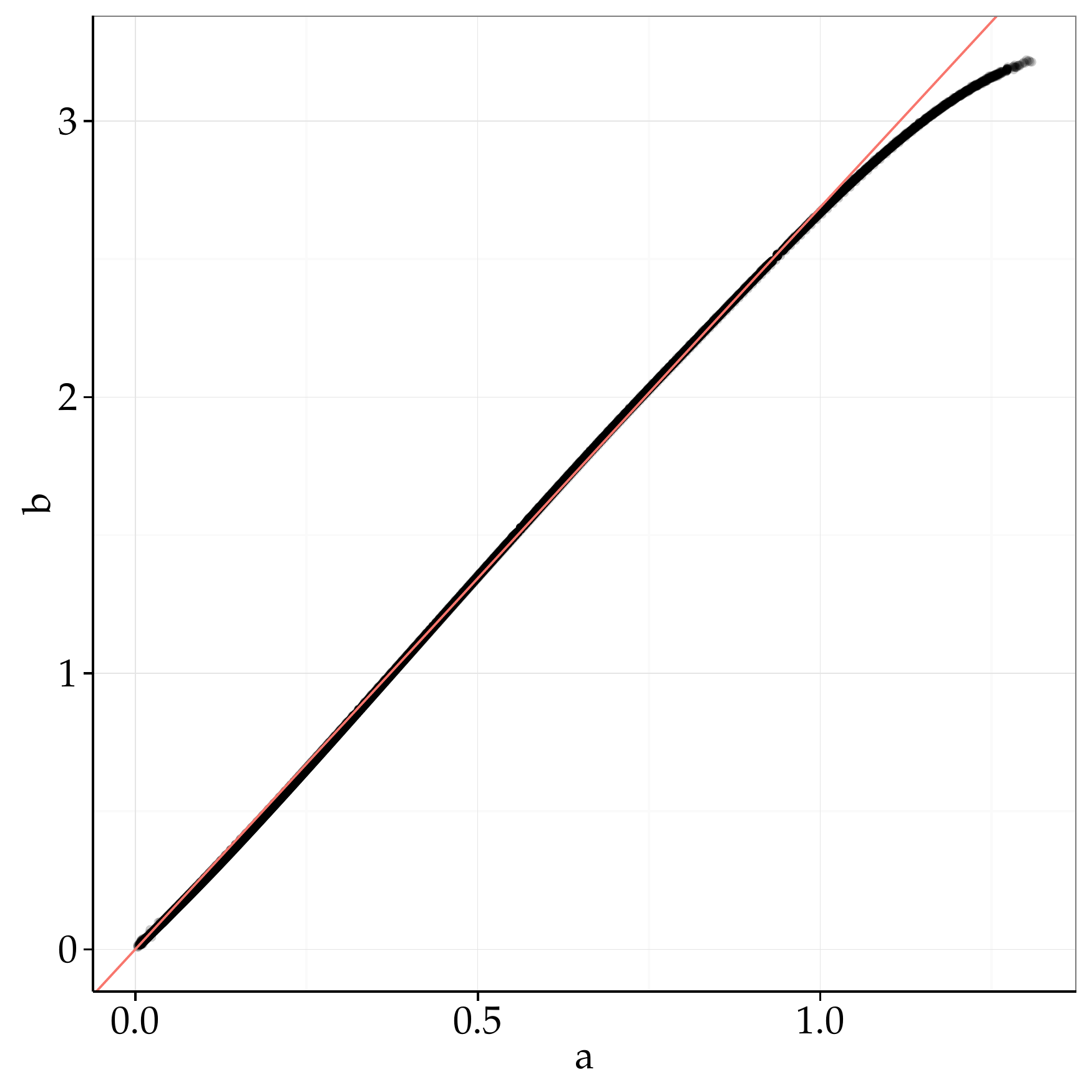}}\protect\caption{Left plot: collapse plot showing the scaling relationship between $a=E_\infty(\hat\theta)$ and $R_0/n$ for 12465 couples $(R_0,n)$, with $100<n<504$; the y axis is in log scale. Right plot: $b$ as a function of $a$, and a linear fit $b\simeq 2.67a$ for $a<1$; same parameters as in the left plot.\label{fig:R0N_vs_a}}
\end{figure}

Let us start with $a=E_\infty(\hat{\theta})$. Left plot of Fig.\ \ref{fig:R0N_vs_a}  shows that $a(R_0,n)=a(R_0/n)$ for $n>100$: the collapse, while not perfect, is remarkable (there are 12645 points in this figure). In other words, $R_0\simeq \gamma  n$ with fixed $\gamma$ at least for $n>100$ and $\theta>0.001$ (a t-statistics of 0.01), as in the large $\theta  n$ limit, although $n\theta=0.1$ in this case is far from being large. Figure \ref{fig:R0N_vs_a} also makes it clear that $1-R_0 /N$ decreases faster than an exponential, which makes sense since it asymptotically follows a Gaussian function (cf. Appendix B). Note that the scaling $R_0\propto n$ assumes that price returns do have a trend. In other words, the Sharpe ratio in the region where $R_0 \propto \sqrt{n}$ will be under-estimated, but they correspond to negligible trends. 

Let us now turn to $b(R_0,n)$. It turns out that there is a linear relationship $b\simeq8/3a$ in the region $a<1$ (see the right plot of Fig.\ \ref{fig:R0N_vs_a}), the collapse being remarkable. This region is  relevant to Finance: for example, if $a=1$, and $n=100$, the t-statistics would be $10$, a rarity. Thus, the whole calibration may rest on the determination of $a(R_0/n)$, since

\begin{equation}\label{eq:estim_theta}
\hat \theta \simeq a(\hat R_0/n)\left[1-\frac{8}{3}\hat\nu^{-3/2}\right]
\end{equation}

As $a$ is a smooth function, we first round $R_0/n$ to a precision of 0.01, and compute the average of $a(r)$ where $r$ is the  rounded  value of $R_0/n$. Finally, a spline is calibrated on this coarse-grained relationship, with the additional the coordinates (0,0) for the sake of convergence for very small values of $R_0/n$. Thus, simple scaling arguments made it possible to build an numerical estimation method for any $n$, $\nu$ and $R_0$ that rests on a single function.

\section{Application to real data}

The i.i.d. assumption is totally unrealistic regarding asset price returns, if only because of volatility heteroskedasticity. Applying straightforwardly the above estimator would therefore make little sense on long time series.  The approach followed here is to consider smaller time windows and to assume that stationarity approximately holds in each time window. The second current limitation of the proposed estimator to keep in mind  here is that it does not account explicitly for skewness. At any rate, this section is meant to provide a clear illustration of how different the estimates of both methods may be.

In order to find the corresponding Sharpe ratio, we assume that price returns are conditionally leptokurtic  \citep{bollerslev1987conditionally}:  in each time window, we fit the returns with Student's t-distribution by maximum likelihood and obtain an estimate $\hat{\nu}$ and use Eq.\  (\ref{eq:estim_theta}).

\begin{figure}
\centerline{\includegraphics[width=0.5\textwidth]{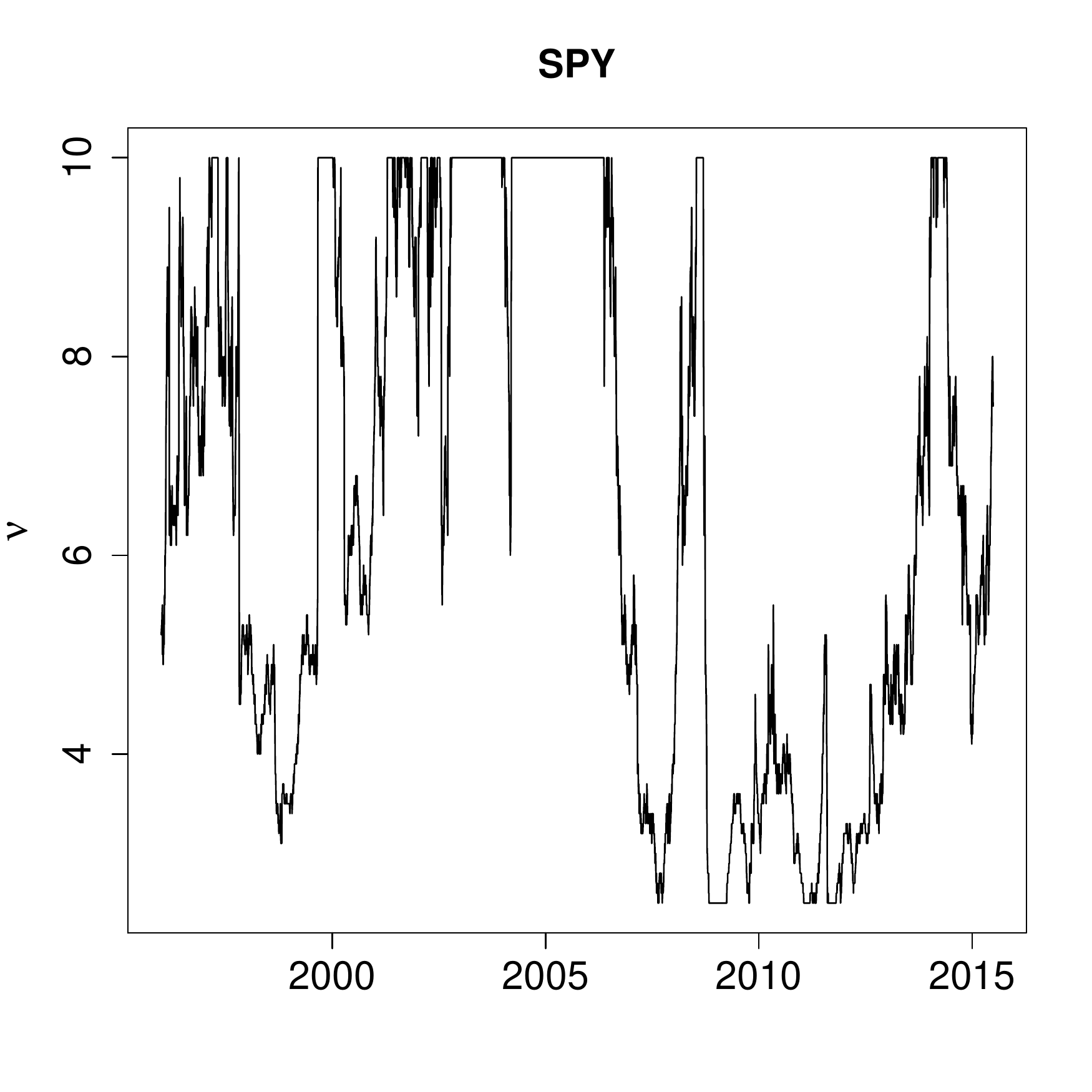}\includegraphics[width=0.5\textwidth]{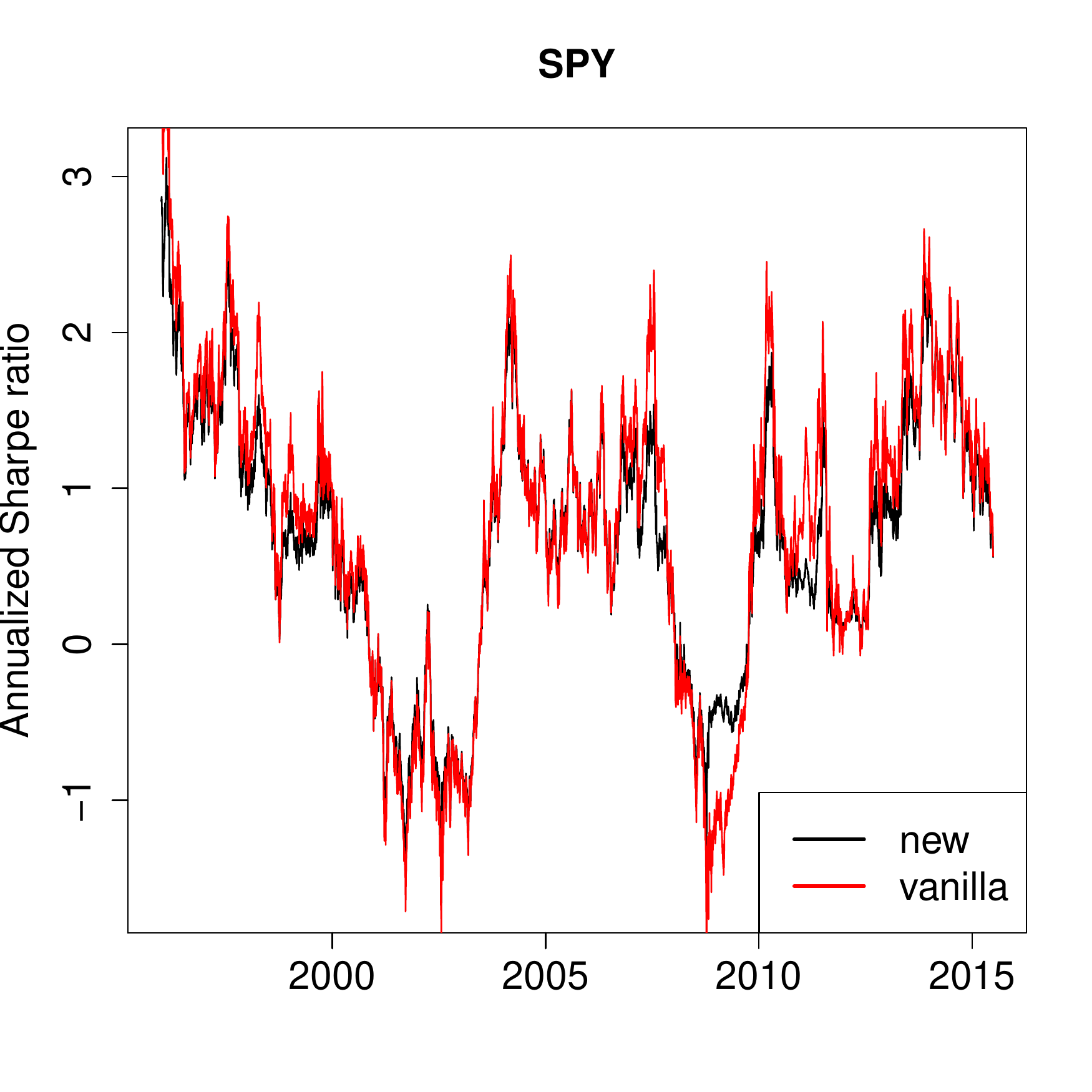}}
\caption{Left plot: parametric fit of the number of degrees of freedom of Student's t-distribtuion in a sliding window of 252 close-to-close returns of SPY. Right plot: estimated Sharpe ratios with the new estimator and from a vanilla estimation. 1000 permutations have been used to estimate $R_0$. \label{fig:nu_SR}}
\end{figure}

\begin{figure}
\centerline{\includegraphics[width=0.5\textwidth]{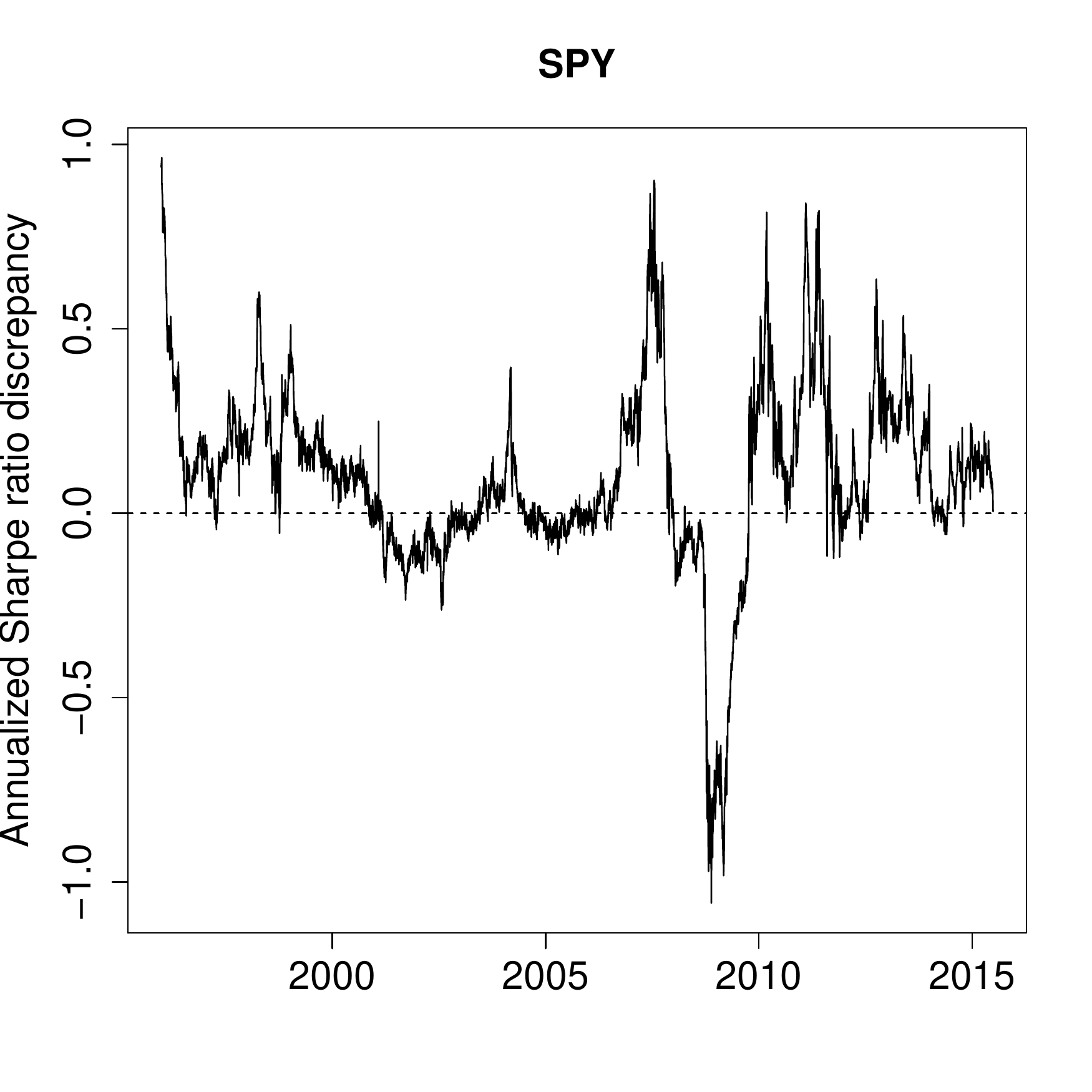}}
\caption{Discrepancy of the estimates of the annualized Sharpe ratio of SPY with moving time windows of 252 days between the new and the vanilla estimators. 1000 permutations have been used to estimate $R_0$.  \label{fig:deltaSR}}
\end{figure}

Figure \ref{fig:nu_SR} shows the difference between annualized Sharpe ratios of SPY estimated with the new and vanilla estimator.  When $\nu$ is larger that 10, both estimators yield almost the same Sharpe ratio, as expected from Eq.\ \ref{eq:fitSR_vs_nu}. On the other hand, when tails are heavier, i.e. when $\nu<10$, the two estimates significantly differ. Indeed, the new term in Eq.\ (\ref{eq:m(n)_student_main}) with respect to Eq.\ (\ref{eq:mn_Gauss}) implies that vanilla estimates are too large in absolute values. This is confirmed in Fig.\ \ref{fig:deltaSR}. The difference between both estimates is very large in leptokurtic times, e.g. in 2008 and 2009; in addition, in these difficult times, the new estimator is clearly less volatile, which is in line with its better efficiency.

\begin{figure}
\centerline{
\includegraphics[width=0.6\textwidth]{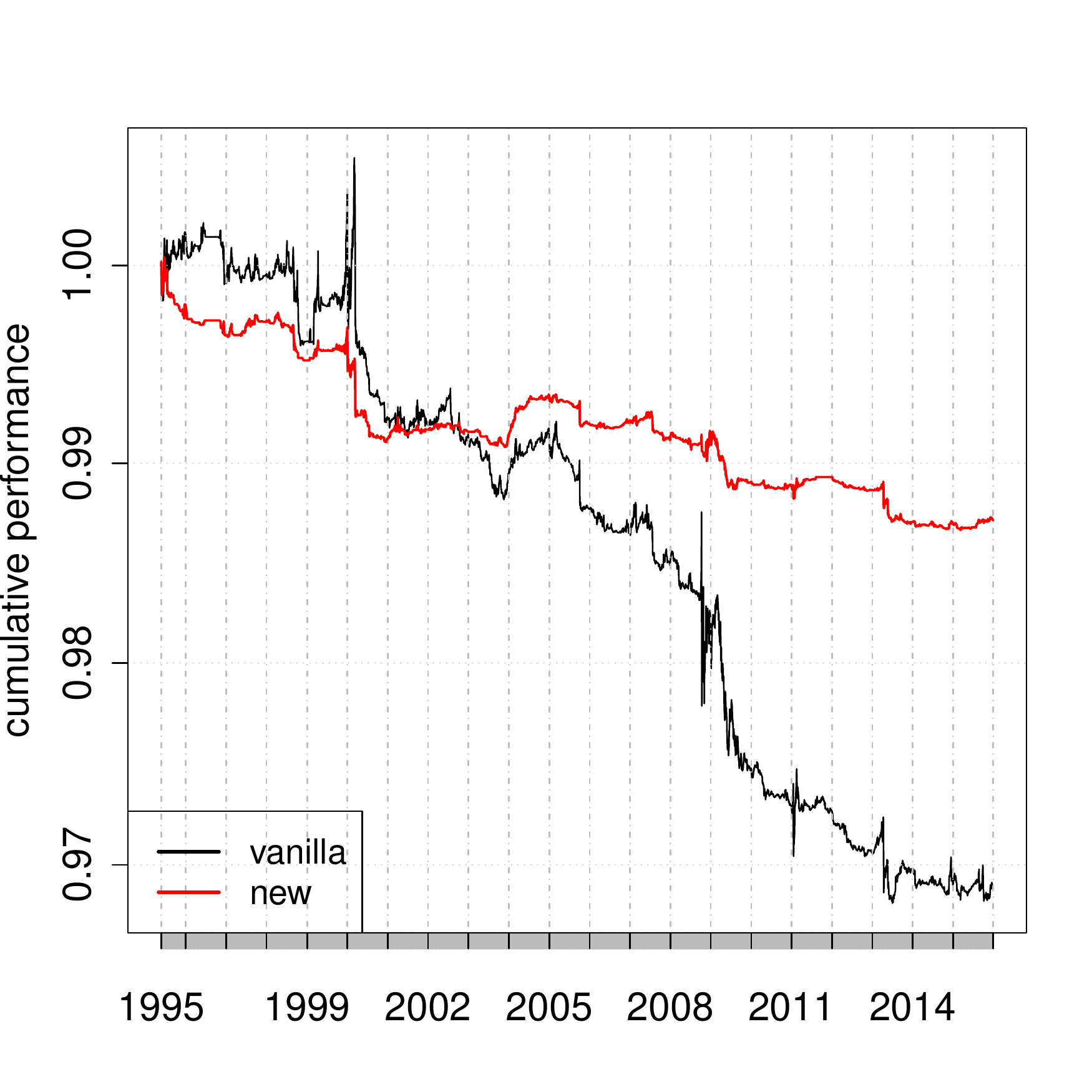}}
\caption{Cumulated performance of a trading strategy consisting in investing when the estimated annualized Sharpe ratio is larger than 1 in absolute value; short positions are allowed in case of negative Sharpe ratio; estimates over rolling windows of 100 trading days (close-to-close price returns). Unbiased historical database of 3449 US liquid equities. 1000 permutations have been used to estimate $R_0$.\label{fig:cumperfs}}
\end{figure}

As a side note, the fact that the moment-based method overestimates the Sharpe ratio (and the t-statistic) in leptokurtic times means that using it for trading purposes leads to taking wrong trading decisions more often (the power of the r-statistics is indeed much larger than that of t-statistics for heavy-tailed data \citep{challet2015rstatistics}). Let us try the following naive trading strategy (without transaction costs): whenever the estimated annualized Sharpe ratio is larger than 1 in absolute value in the last 100 close-to-close price returns, one takes a long or short single-day position, depending on the sign of the Sharpe ratio (with a one day lag). We use an unbiased historical data set of US equities (1995-01-01 to 2015-06-30) and focus on liquid assets, i.e. whose price is larger than \$20 and 60-day rolling median daily volume is larger than 250000 shares.  Figure \ref{fig:cumperfs} reports the cumulated performance of this strategy when applied to all 3449 US equities for the period . The difference of performance between the two methods is marked in times of large fluctuations (e.g. 2008). Note that the y-axis of this plot is logarithmic so as to avoid fooling the reader \citep{mclean2011fooled}; in addition, it should be noted that the out-of-sample performances plotted in Fig. \ref{fig:cumperfs} is the result of 3449 decisions at each time step, i.e., very many decisions. As a consequence, the origin of the difference of performance between the new and vanilla methods is the relative power of the related statistical tests \citep{challet2015rstatistics}, not an erroneous way of computing compounded returns.

\subsection{Ranking assets}
\label{ranking}
It is worth discussing the relevance and practical value of the proposed Sharpe ratio estimator beyond its much improved precision in leptokurtic times and the appreciable fact that it is unbiased. The industry is not only interested in the actual Sharpe ratio values of a group of assets (stocks, hedge funds, etc.), but also in ranking them. Thus, an important question is to assess whether the new method ranks assets in a different way than vanilla Sharpe ratio estimation. If this is clearly not the case, by extension, the new method brings a valuable alternative way to rank assets.

Two preliminary remarks. First, the fact that both methods estimate the same quantity, but that one method is clearly much more efficient for non-Gaussian variables implies that the correlation of the asset ranks cannot be 1 because of the greater fluctuations of the vanilla estimator. Second, as explained above, the Sharpe ratio corresponding to an estimated $R_0$ depends on the tail index of the Student distribution. Since the new method is unbiased and the vanilla one is biased for heavy-tailed distribution, and since the estimated tail exponents at any given time will vary from asset to asset, one cannot expect the two methods to yield on average equivalent ranking, even asymptotically. In other words, the location-shape argument of \cite{schuhmacher2011sufficient} does not hold for assets with heterogeneous tail indices, as noted e.g. in \cite{zakamulin2010choice}. 

Let me take an example: looking once again at Fig.\ \ref{fig:SR_vs_nu} makes it clear that the ranking of $R_0$, i.e., of the vanilla estimation method, may not be the same one as the ranking according to the new method. Assume that asset 1 has $R_{0,1}=40$ and asset 2 $R_{0,2}=50$; neglecting the fact that $\nu_1<\infty$ and $\nu_2<\infty$ is quite possible, the vanilla method attributes a better rank to asset 2. Now, say $\nu_1=10$; as soon as $\nu_2<4.4$, asset 1 must be attributed a better rank than asset 2 (provided that the estimations of the tail exponents are precise enough). \label{par:discussion_nu}

 All the above shows the crucial role of $\nu$ and sheds a new light on the debate on whether all risk measures are equivalent asymptotically or not. The point is that many of them may be biased for non-Gaussian variables in an equivalent way.  For example,  \cite{auer2013robust} find that the rankings of hedge fund performance with the largest Sharpe ratios are most stable. But its results rest on measures that overestimate Sharpe ratios for heavy-tailed returns; thus, as sizeable fraction of these large Sharpe ratios may simply be the results of the methods' biases.

\begin{figure}
\centering{\includegraphics*[width=0.5\textwidth]{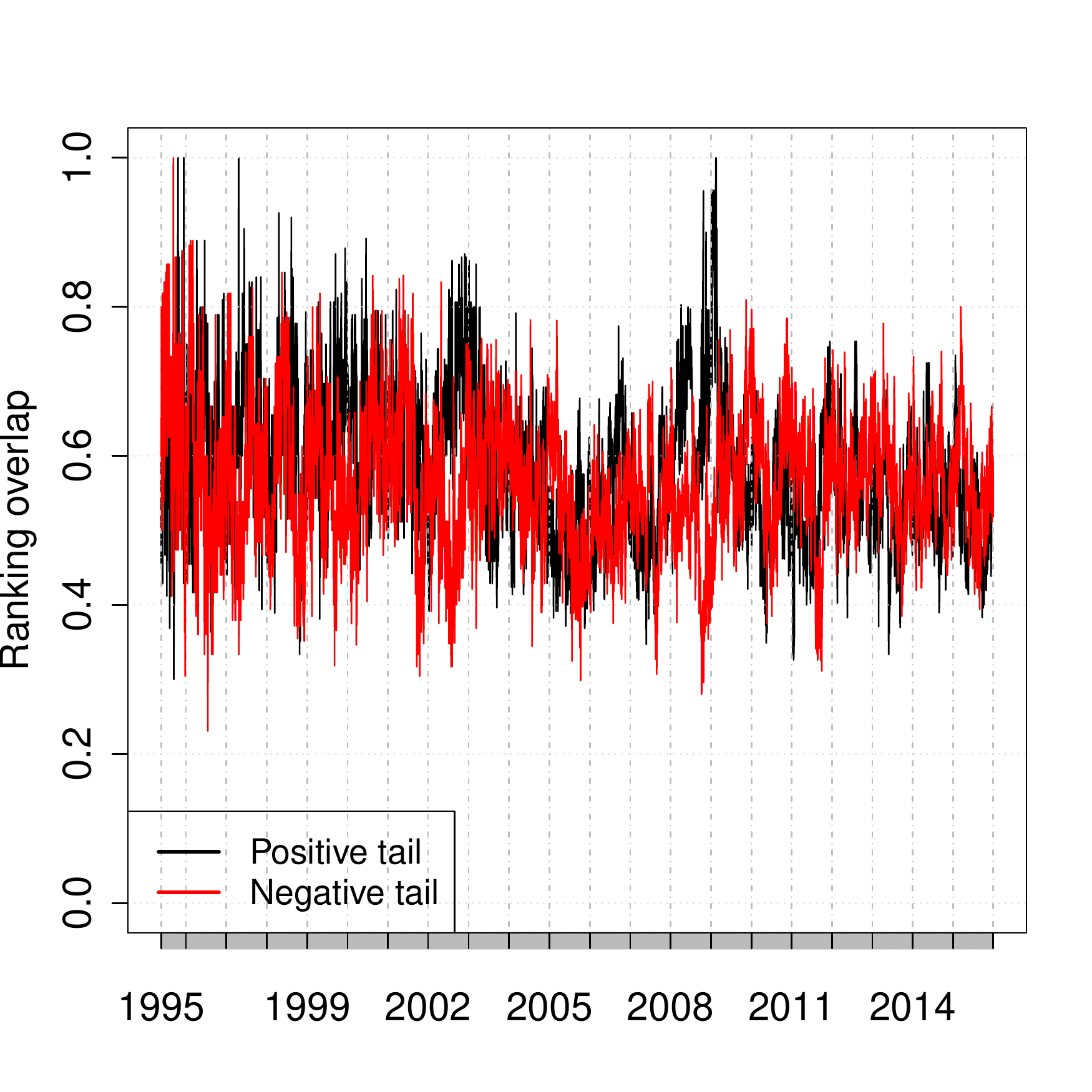}~~~\includegraphics*[width=0.5\textwidth]{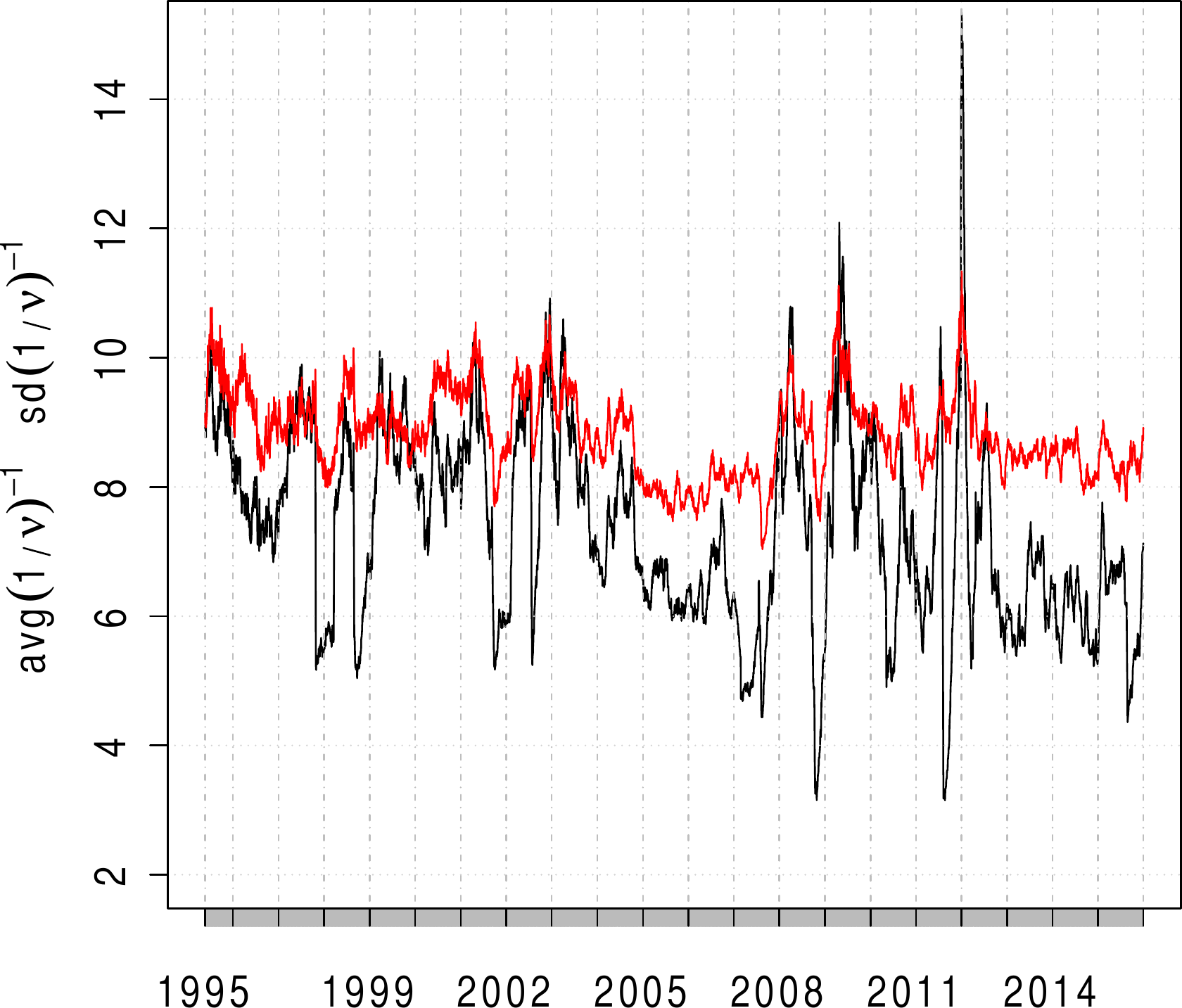}\protect\caption{Left: evolution of the fraction of common assets between rankings for the moment-free and vanilla Sharpe ration estimation methods; black lines: top 5\% positive ratios, red lines: smallest 5\% ratios; calibration over 100 days; unbiased historical database of 3449 liquid US equities. Right: effective measure of average (red lines) and standard deviation (black lines) of tail exponents as a function of time for the 3449 US equities.}
\label{fig:frac_SR_posneg}}
\end{figure}

Let me first focus on the 5\% best and worst estimated Sharpe ratios.  Figure \ref{fig:frac_SR_posneg} (left plot) shows that the fraction of common assets in these centiles is significantly different from 1, except notably for positive Sharpe ratios in 2008. As expected, this fraction decreases when the heterogeneity of tail exponents increases (right plot).

\begin{figure}
\centering{}\includegraphics[width=0.5\textwidth]{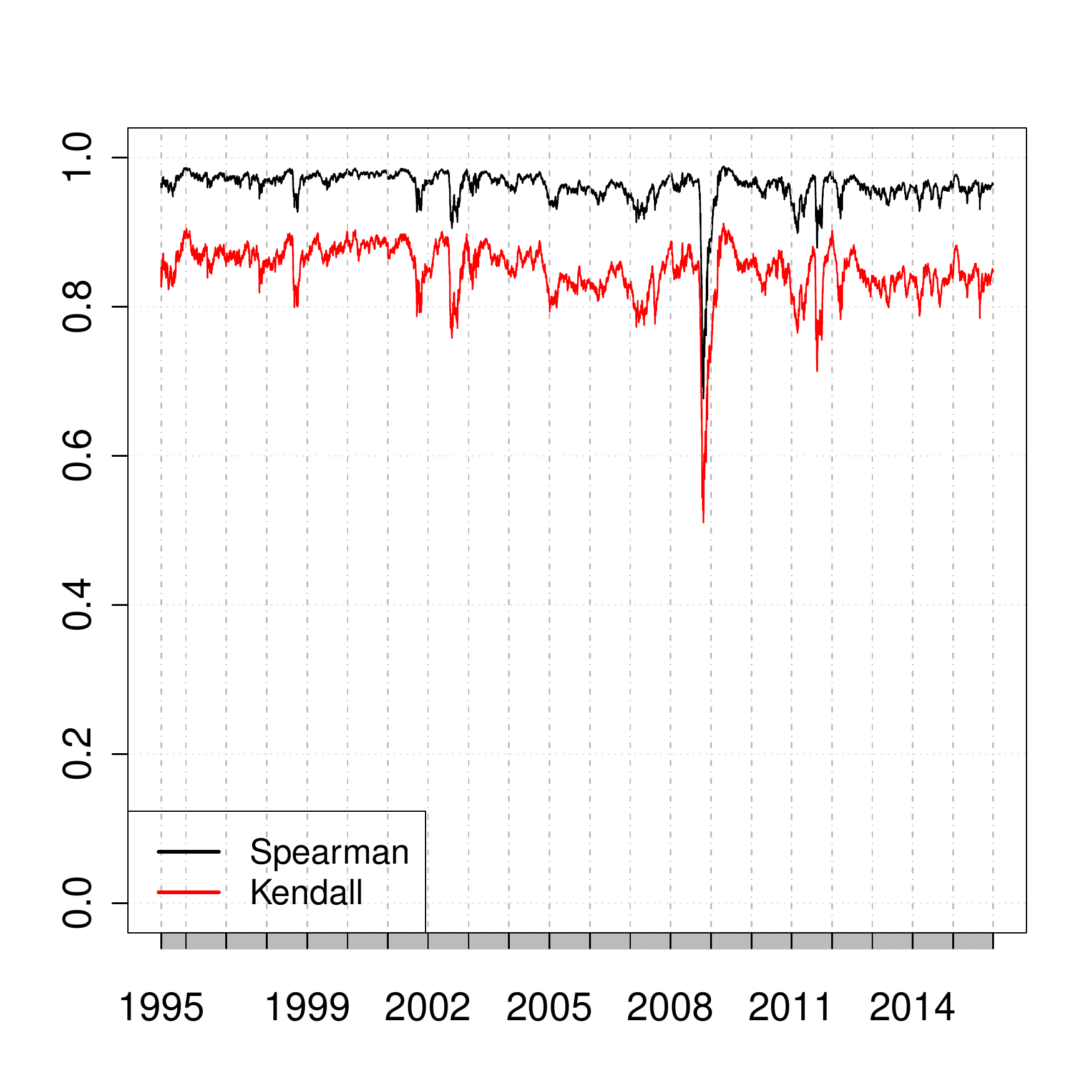}\includegraphics[width=0.5\textwidth]{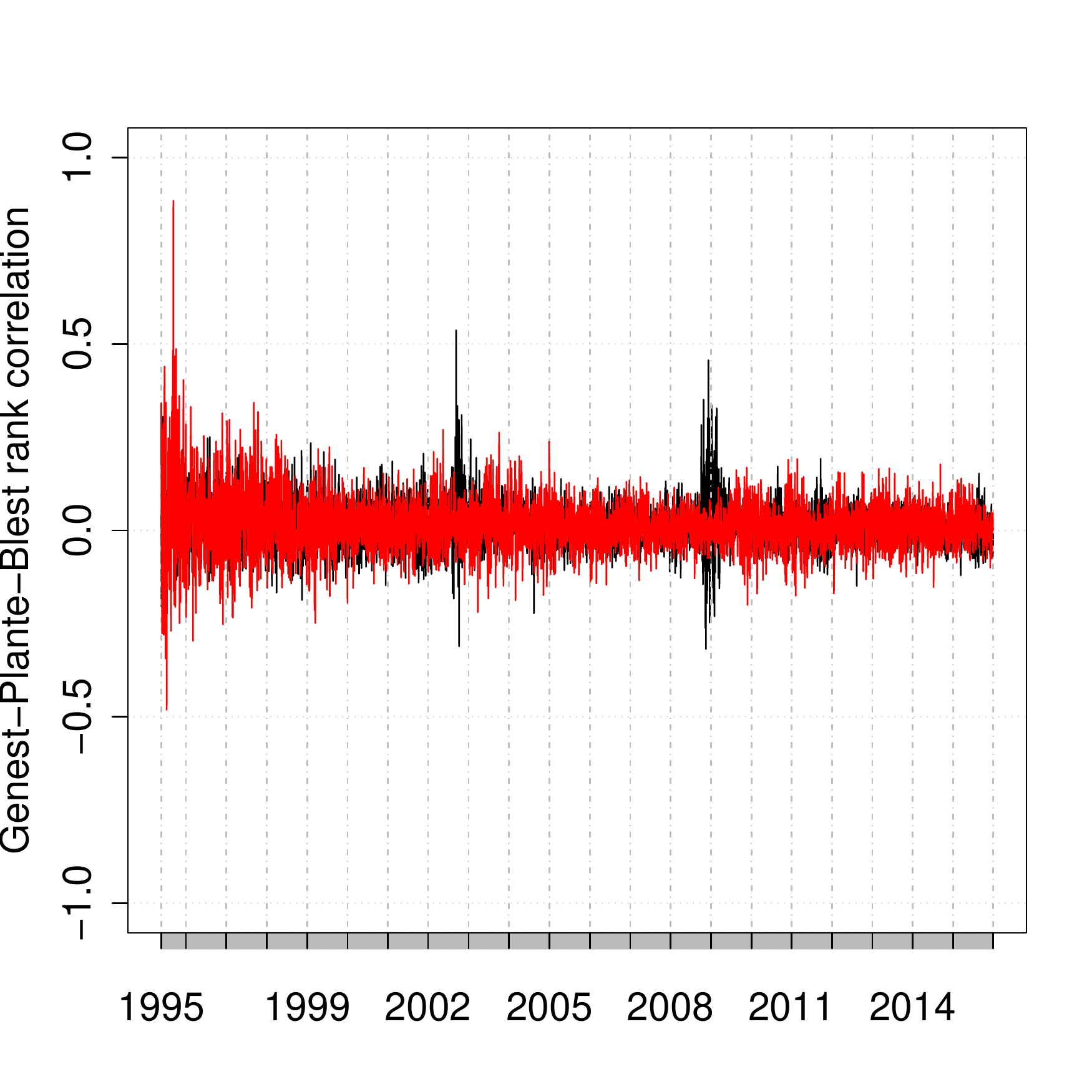}\protect\caption{ Spearman, Kendall (left plot) and Genest-Plante-Blast (right plot) rank correlations between the ranks of the Sharpe ratios obtained by record statistics and the usual method as a function of time (black lines: positive Sharpe ratios, red lines: negative Sharpe ratios). Sliding calibration windows of 100 days.   \label{fig:spearman_rank_SNR0}}
\end{figure}

As expected, rankings differ more for assets with a small $\nu$,  i.e., with heavy tails. Left plot of Fig.\ \ref{fig:spearman_rank_SNR0} reports the time evolution of Spearman and Kendall rank correlation of all the assets for both methods. As in \cite{auer2013robust}, an alternative rank correlation measure \citep{blest2000corr,genest2003blest}, more sensitive to the rank of the largest values of data, is displayed in the right hand side of this figure; it is slightly above zero on average, with large fluctuations in 2003 and 2009, which echoes the decrease of both Spearman and Kendall correlations at those dates, but not at other dates. Thus, ranking equities with usual moment-based methods or the new moment-free method may yield very significantly different results in the case of daily price return of equities. Hedge fund performance returns have a monthly resolution and are thus much closer to Gaussian variables, owing to the central limit theorem (see e.g. \cite{bouchaudpotters}), which may also explain why previous studies did not find striking differences of ranking for most performance measures.

\section{Discussion}

The proposed Sharpe ratio estimator is robust, efficient, and well-behaved as it does not rely on moment estimation. Large returns are not
regarded as outliers, but contribute to record statistics in a smooth
way. In addition, a real outlier (due e.g. to a data error, or a neglected corporate action) may only create a single spurious additional price record, while two outliers of the same magnitude and opposite signs have only a mild influence on $R_0$.  Finally, the robustness of the estimator lies in the fact that the latter is based only on the duration of drawdowns, not on their amplitudes. This is to be contrasted with other quantities related to drawdowns. For example the expectation of the maximum drawdown of a Brownian motion is a known function of the Sharpe ratio \citep{magdon2003maximum}, but is very sensitive to outliers by definition.

Because of the lack of exact results, using this estimator requires for the time being numerical calibration, which has been much simplified by scaling arguments. Estimating Sharpe ratios with price record/drawdown statistics is not
limited to Student's t-distributed returns, as indeed one may
calibrate their relationships for any return distribution with finite
variance. In addition, the method introduced in this paper provides a
generic way to build many types of estimators with record statistics
as long as the relationship between price record statistics and the
measurable to estimate is monotonic. For example, it may be used to
estimate the drift of a L\'evy process.

The main limitations of the proposed estimator are the assumptions of
i.i.d and symmetric increments. Both can be accounted for numerically
for the time being. An interesting challenge is to incorporate serial
correlations into the analytical computation of record statistics:
numerical results point to simple corrections in the case of AR(1) and
GARCH(1,1) models \citep{wergen2014modelingrecordstock}.  Practically,
a way to respect return auto-correlation and volatility
heteroskedasticity is to use a kind of block-bootstraps, as in
\citet{ledoit2008robust}.

An R package entitled {\tt sharpeRratio} implements the new estimator and is available on CRAN at {\url{https://cran.r-project.org/web/packages/sharpeRratio/index.html}}; a Python version is available at PyPi {\url{https://pypi.python.org/pypi/pysharperratio/0.1.10}}. 

An interactive webpage which reproduces the  plots of Section 3 for any asset symbol and time period may be found at\\ {\url{https://brillant.shinyapps.io/moment-free_Sharpe_ratio}}\ .

\appendix

\section{Expected number of records in the vanishing Sharpe ratio limit for Student's t-distributed price returns}

In this  limit, one may use a first order expansion
of the reciprocal cumulative function 
\begin{equation}
P(S_{n}>0)=\frac{1}{2}+P^{(n)}(0)cn+O([cn]^{2}).\label{eq:pk>(x)}
\end{equation}

One therefore needs to compute $P^{(n)}(0)$. Since the increments
are assumed to be independent, 
\[
P^{(n)}(0)=\frac{1}{2\pi}\int_{-\infty}^{\infty}\phi^{(n)}(t)dt=\frac{1}{2\pi}\int_{-\infty}^{\infty}[\phi(t)]^{n}dt,
\]
where $\phi^{(n)}(t)$ is the characteristic function of $P^{(n)}(x)$,
and $\phi(t)$ that of $P^{(1)}(r)=P(r)$. Equation~\eqref{eq:pk>(x)}
requires the computation of $P^{(n)}(0)$, which is impossible for any $n$ and $\nu$. However
the $\nu=3$ case leads to workable expressions. One finds $P^{(n)}(0)=\frac{e^{n}}{\sigma\pi}E{}_{-n}(n)$,
where $E_{n}(z)$ is the exponential integral function. The specific
form $n=-z$ of the exponential integral function is easy to compute
in a recursive way by integration by parts:
\begin{eqnarray*}
E_{-k-n}(k) & = & \frac{e^{-k}}{k}+\frac{k-n}{k}E_{-(k-n-1)}(k)\\
E_{0}(k) & = & \frac{e^{-k}}{k}.
\end{eqnarray*}

Therefore, after some elementary computations, $E_{-n}(n)=\frac{e^{-n}}{n}\frac{n!}{n^{n}}\left(\sum_{s=0}^{n}\frac{n^{s}}{s!}\right)$
and

\begin{equation}
P^{(n)}(0)=\frac{1}{\sigma\pi}\,\frac{1}{n}\,\frac{n!}{n^{n}}\sum_{s=0}^{n}\frac{n^{s}}{s!}.\label{eq:p_k(0)}
\end{equation}

Using the asymptotic expansion $K_{n}=\sum_{s=0}^{n}\frac{n^{s}}{s!}=e^{n}[\frac{1}{2}+\sqrt{\frac{2}{3\pi}}\frac{1}{\sqrt{n}}+O(n^{-1})]$
and the usual Stirling expansion, Eq.~\eqref{eq:p_k(0)} 
becomes $P^{(n)}(0)=\frac{1}{\sigma}\frac{1}{\sqrt{2\pi n}}+\frac{2}{\sigma\pi \sqrt{3}n}+O(n^{-3/2})$
and thus, in the case of small drifts, Eq.~\eqref{eq:pk>(x)} reads
\[
P(S_{n}>0)=\frac{1}{2}+\frac{c}{\sigma}\frac{2}{\pi\sqrt{3}}+\frac{c}{\sigma}\sqrt{\frac{n}{2\pi}}+O(n^{-1/2}).
\]

Higher-order expansions of $K_{n}$ and $n!$ contribute terms of
order $n^{-1/2}$ that are negligible.  It is noteworthy that the additional correction for Student increments
does not depend on $n$; accordingly, it is relevant for any value
of $n$ and has a larger relative weight for smaller $n$; this is
consistent with the fact that convolutions of Student's t-distributions
with $\nu=3$ converge to a Gaussian distribution. Sparre Andersen
theorem yields 

\begin{equation}
\tilde{q}_{-}(z)=\frac{1}{\sqrt{1-z}}\left(1+\sum_{n=1}^{\infty}\frac{c}{\sigma}\frac{z^{n}}{\sqrt{2\pi n}}\right)-\frac{c}{\sigma}\frac{2}{\pi\sqrt{3}}\frac{\log(1-z)}{\sqrt{1-z}}+O[(c/\sigma)^{2}].\label{eq:tilde q(z)}
\end{equation}

The generating function of the number of records is then \citep{ledoussal2009driven}
\[
\tilde{m}_{+}(z)\simeq\frac{1}{(1-z)^{3/2}}\left[1+\frac{c}{\sqrt{2\pi}\sigma}\sum_{n=1}^{\infty}\frac{z^{n}}{\sqrt{n}}-\frac{2c}{\sigma\pi\sqrt{3}}\log(1-z)\right].
\]

The two first terms in the brackets are the same ones as those of
Gaussian biased random walks \citep{majumdar2012record}. The third term is new and due to the difference between a Gaussian and a t-distribution at the origin. Whereas until this point the approximations are controlled, current literature on this topic goes one step further: asymptotic results are (very) roughly obtained by approximating divergent partial sums. This captures the way the sum diverges, without much control over the precision of the prefactor. At any rate, as the prefactor is not essential to our purpose, we have followed the same route, which yields
\begin{equation}
-\frac{1}{(1-z)^{3/2}}\log(1-z)\simeq\frac{2}{\sqrt{\pi}}\sum_{n\ge1}\left[2\sqrt{n}\left(\mbox{atanh}\sqrt{1-\frac{1}{n}}-\sqrt{1-\frac{1}{n}}\right)\right]z^{n},\label{eq:approx_sum_int}
\end{equation}

 which is not a very good approximation even for large $n$ but gives the correct asymptotic $\sqrt{n}$ dependence, with an additional logarithmic correction brought by $\mbox{atanh}\sqrt{1-\frac{1}{n}}-\sqrt{1-\frac{1}{n}}$. Finally, approximating $n$ by $n-1$ as in \cite{wergen2011record} and identifying each term of the generating function with the value of $n$ one is interested in gives
\begin{eqnarray}
E(R_+)(c/\sigma,n)&\simeq&\frac{2\sqrt{n}}{\sqrt{\pi}}+\frac{c\sqrt{2}}{\sigma\pi}\left[n\arctan(\sqrt{n})-\sqrt{n}\right]\\\nonumber&&+\frac{c}{\sigma}\frac{8}{\sqrt{3}\pi^{3/2}}\sqrt{n}\left(\mbox{atanh}\sqrt{1-\frac{1}{n}}-\sqrt{1-\frac{1}{n}}\right)\nonumber.\label{eq:m(n)_student}
\end{eqnarray}

Given its derivation, this formula is relevant in the limit $cn\ll\sigma$
and large $n$, or equivalently $c/\sigma\sqrt{n}\ll1/\sqrt{n}$, i.e., in the limit of vanishingly small Sharpe ratios.

\section{Expected number of price records in the large $cn/\sigma$ limit}

The  $cn/\sigma\gg 1$, $n\gg1$ limit
also makes it possible to derive some analytical insights. \citet{majumdar2012record}
give results for large, but not too large,  $cn/\sigma$. Indeed,
the central limit theorem states that the convergence of the distribution
of convoluted variables to a Gaussian distribution occurs from the center of the distribution.
This implies that the tails of any non-Gaussian distribution are non-Gaussian.
Thus, intuitively, when $cn/\sigma$ is large enough (whose meaning will be
discussed below), $P(x_n<cn)$ comes from the non-Gaussian tails. This
will lead to markedly different results for Student's t-distributions
since the tails of convoluted t-distributions keep their power-law
nature. \citet{bouchaudpotters} give an intuitive argument to compute
the $n$-time convoluted return $r_{0}^{(n)}$ at which the Student and Gaussian parts of the
distribution have equal importance and find that $r_0^{(n)}\simeq\sigma\sqrt{n\log n}$
for $\nu=3$. This means that the value of $n_{0}$ at which the power-law
tail starts to prevail is such that $cn_{0}\simeq\sigma\sqrt{n_{0}\log n_{0}}$,
i.e., 
\begin{equation}
\frac{c}{\sigma}\sqrt{n_{0}}\simeq\sqrt{\log n_{0}}.\label{eq:n0_vs_cs}
\end{equation}

\begin{figure}

\begin{centering}
\includegraphics[width=0.6\textwidth]{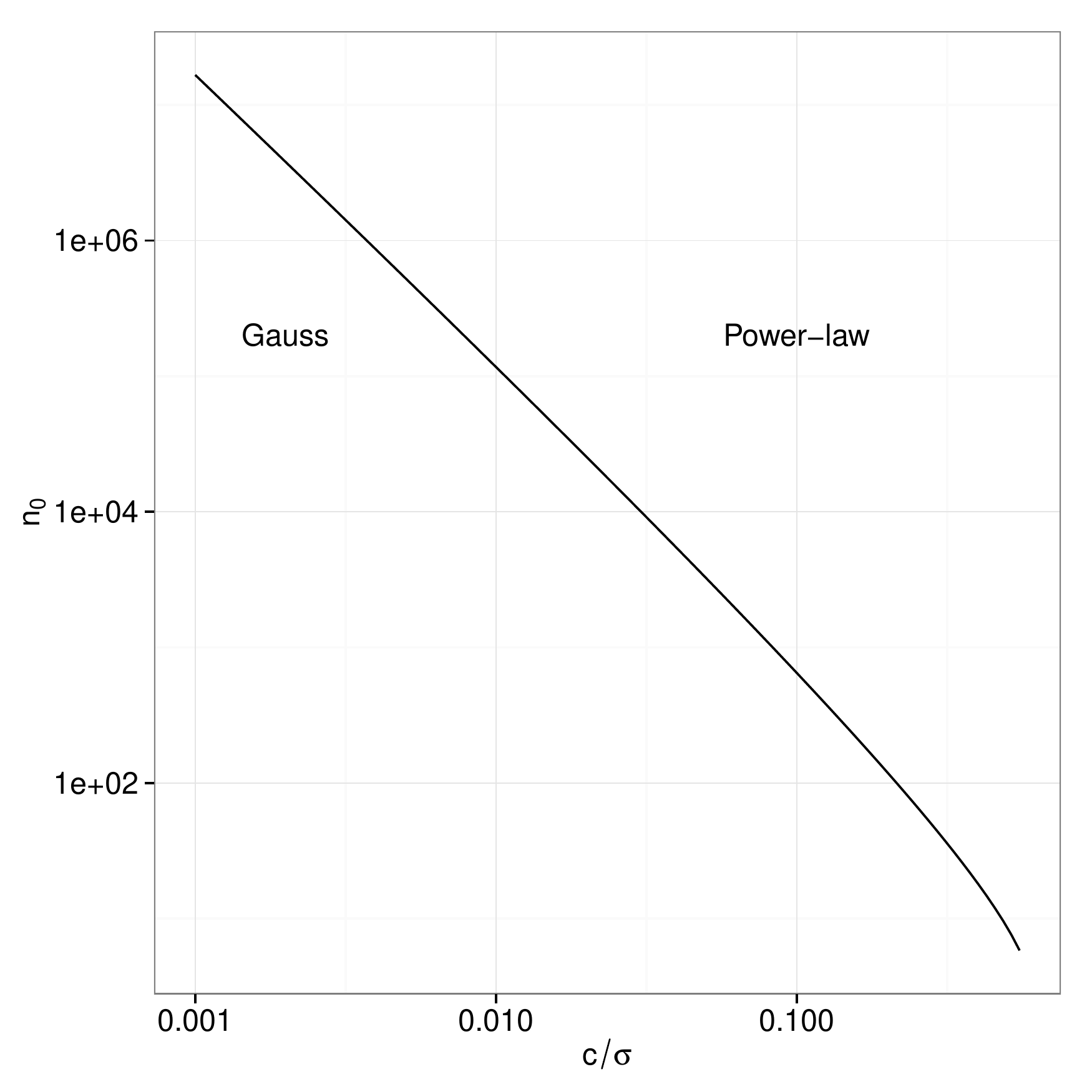}\protect\caption{Limiting $n_{0}$ as a function of $c/\sigma$, from Eq.~\eqref{eq:n0_vs_cs}.
Convoluted Student's t-distributions may be approximated by a Gaussian
distribution below the continuous line, and by a power-law above this
line. \label{fig:st_vs_gauss}}

\par\end{centering}

\end{figure}
Since the convoluted distribution has a continuous first derivative,
there is no sharp transition between the Gaussian and power-law regimes,
hence $n_{0}$ only approximately indicates where the Gaussian approximation
begins to break down. Figure \ref{fig:st_vs_gauss} plots $n_{0}(c/\sigma)$
and shows these two regions. In the region well below the line,
a Gaussian approximation holds for Student convolutions. Reversely,
when $n\gg n_{0}(c/\sigma)$, 
\[
P(S_{n}<0)\simeq\int_{cn}^{\infty}\frac{2\sigma^{3}n}{\pi x^{4}}dx=\left(\frac{\sigma}{c}\right)^{3}\frac{2}{3\pi n^{2}},
\]

hence $\log\left(\tilde{q}_{-}(z)\right)=\left(\frac{\sigma}{c}\right)^{3}\frac{6\sqrt{3}}{\pi}\sum\limits _{n=1}^{\infty}\frac{z^{n}}{n^{3}}$,
thus 
\[
\tilde{m}_{+}(z)=\frac{1}{(1-z)^{2}}\exp\left[-\left(\frac{\sigma}{c}\right)^{3}\frac{2}{3\pi}\sum\limits _{n=1}^{\infty}\frac{z^{n}}{n^{3}}\right]\simeq\frac{1}{(1-z)^{2}}\left[1-\left(\frac{\sigma}{c}\right)^{3}\frac{2}{3\pi}\sum\limits _{n=1}^{\infty}\frac{z^{n}}{n^{3}}+O\left[\left(\frac{c}{\sigma}\right)^{6}\right]\right].
\]

Finally, one finds without major difficulty

\[
\tilde{m}_{+}(z)\simeq\sum_{n\ge0}\left[(n+1)\left(1-\left(\frac{\sigma}{c}\right)^{3}\frac{2}{3\pi}K[1+O(n^{-1})\right)\right]z^{n}
\]

and

\[
m_{+}(n)\simeq n\left[1-\left(\frac{\sigma}{c}\right)^{3}\frac{2}{3\pi}K\right].
\]

\begin{figure}
\centering{}\includegraphics[width=0.6\textwidth]{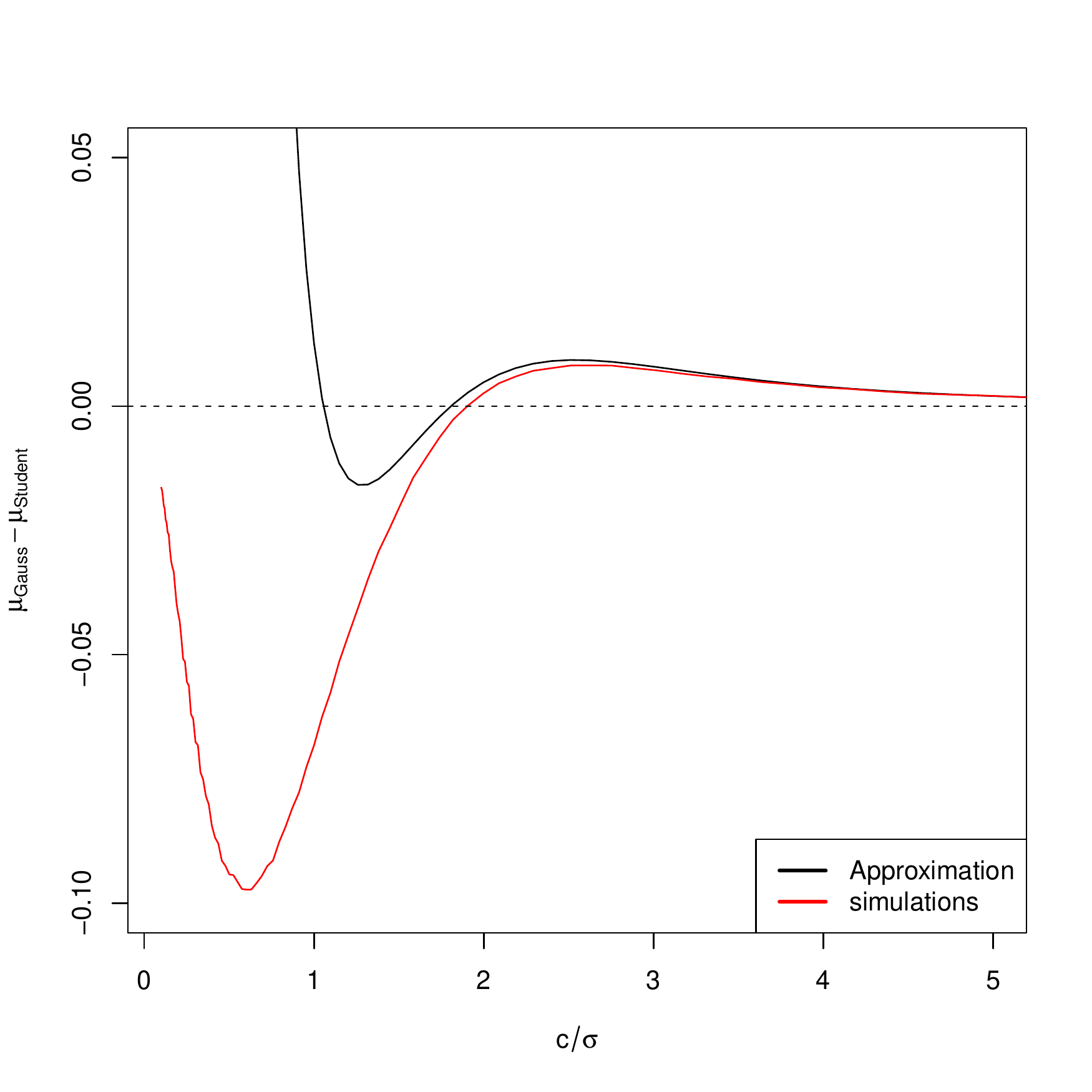}\protect\caption{Large signal-to-noise ratio limit: difference of record rate between
random walks with Gaussian increments and Student's t-distributed
increments. The rate was determined as the average slope $[m(200)-m(100)]/100$
(averages over $10^{5}$ samples).\label{fig:Large-signal-to-noise-ratio}}
\end{figure}

Numerically, $K\simeq1.202\simeq\frac{6}{5}$ for large $n$; approximating sums with
integrals yields the very different $K=1/2$. Thus the number of records
increases linearly for large $n$ $m_{+}(n)\simeq n\mu_{Student}$
with an asymptotic rate given by $\mu_{Student}\simeq1-\left(\frac{\sigma}{c}\right)^{3}\frac{4}{5\pi}$,
to be compared with $\mu_{Gauss}\simeq1-\frac{\sigma}{c}\frac{1}{\sqrt{2\pi}}e^{-\frac{c^{2}}{2\sigma^{2}}}$.
Figure \ref{fig:Large-signal-to-noise-ratio} plots the difference
of the record rate between Gaussian- and Student's t-distributed ($\nu=3$)
increments as a function of $c/\sigma$. Whereas 
the number of records of random walks with Student increments are
larger than those with Gaussian ones for small Sharpe ratios, Fig.\ \ref{fig:Large-signal-to-noise-ratio}
shows, somewhat surprisingly, that Gaussian increments lead to a larger
rate of records for very large Sharpe ratios. 

\bibliographystyle{rAMF}
\bibliography{biblio}

\end{document}